\documentstyle[12pt,aasms4,psfig]{article}

\lefthead{K. Krisciunas {\em et al.}}
\righthead{Color Relations for Type Ia Supernovae}

\begin{document}

\title{Uniformity of V {\em minus} Near Infrared
Color Evolution of Type Ia Supernovae,
and Implications for Host Galaxy Extinction Determination}
\author{Kevin Krisciunas$^1$,
N. C. Hastings$^2$,
Karen Loomis$^2$, 
Russet McMillan$^2$,
Armin Rest$^1$,
Adam G. Riess$^3$,
and Christopher Stubbs$^1$}
\affil{$^1$Department of Astronomy, University of Washington, Box 351580,
Seattle, WA 98195$-$1580 \\
$^2$Apache Point Observatory, Astrophysical Research Consortium,
2001 Apache Point Road, P. O. Box 59, Sunspot, NM 88349-0059  \\
$^3$Space Telescope Science Institute, 3700 San Martin Drive,
Baltimore, MD 21218}
\begin {center}
Electronic mail: kevin, hastings, rest, stubbs@astro.washington.edu \\
karen, rmcmillan@apo.nmsu.edu \\
ariess@stsci.edu 
\end {center}

\begin{abstract} 

From an analysis of SNe 1972E, 1980N, 1981B, 1981D, 1983R, 1998bu, 1999cl,
and 1999cp we find that the intrinsic V$-$K colors of Type Ia SNe of
typical luminosity (i.e. with multi-color light curve shape (MLCS)
parameter $-0.4 \lesssim \Delta \lesssim +0.2$ mag) suggest a uniform
color curve. V$-$K colors become bluer linearly with time from roughly one
week before B-band maximum until one week after maximum, after which they
redden linearly until four weeks after maximum. V$-$H colors exhibit very
similar color evolution.  V$-$J colors exhibit slightly more complex
evolution, with greater scatter. The existence of V {\em minus} near
infrared color relations allows the construction of near infrared light
curve templates that are an improvement on those of Elias et al. (1985).

We provide optical BVRI and infrared JHK photometry of the Type Ia
supernovae 1999aa, 1999cl, and 1999cp.  SN 1999aa is an overluminous ``slow
decliner'' (with $\Delta$ = $-$0.47 mag).  SN 1999cp is a moderately bright
SN unreddened in its host. SN 1999cl is extremely reddened in its host. The
V {\em minus} near infrared colors of SN 1999cl yield A$_V$ = 2.01 $\pm$
0.11 mag. This leads to a distance for its host galaxy (M 88) in agreement
with other distance measurements for members of the Virgo cluster.

\end{abstract}


\section{Introduction}

Type Ia supernovae (SNe, singular SN) have become recognized as one of the
most significant standardizable candles for the determination of
extragalactic distances (Suntzeff 2000, Leibundgut 2000, Burrows 2000). As
was the case with Cepheids and RR Lyrae stars, much effort must be
expended to calibrate any new standard candle.  Even while we debate an
{\em absolute} calibration, standardizable candles can be used as
excellent {\em relative} distance indicators; this has been done for some
time with Type Ia supernovae, and since Phillips (1993) identified the
absolute magnitude-decline rate relation, observations of Type Ia SNe have
progressed with a greater sense of direction.

Since absolute distance determinations are a primary goal, let us begin
with the basics. In order to obtain the luminosity distance $d$ (in pc) to
a celestial object from its brightness we must know three things: (1) an
accurate apparent magnitude m of the object; (2) an estimate of the
object's absolute magnitude M; and (3) an estimate of any extinction along
the line of sight.  These four parameters are related by the standard
equation:

\begin {equation}
\rm{M}_{\lambda} \; = \; m_{\lambda} - \; \rm{A}_{\lambda} \; + \; 5 \; - \; 5 \; \rm{log} \; d  ,
\end {equation}

\parindent = 0 mm

where A$_{\lambda}$ is the extinction in magnitudes.  Each 0.1 mag
error due to extinction, absolute magnitude, or apparent magnitude
corresponds to a 5 percent error in the distance to the object.  Thus,
extinction is potentially a serious source of systematic errors in
determining distances.  The primary focus of this paper is to outline a
method of determining the V-band extinction, A$_V$, towards Type Ia
supernovae. For a discussion of the absolute magnitudes see Meikle
(2000).

\parindent = 9 mm

As is well known, the extinction due to dust is an order of magnitude less
in the infrared compared to optical wavelengths.  While it is not new that
a combination of optical and infrared data can be used to investigate
extinction (Sneden et al. 1978), this is not yet routine for supernovae.

\subsection{Apparent magnitudes}

At present more than 100 supernovae are discovered each
year.\footnote[4]{See 
{\tt http://cfa-www.harvard.edu/iau/lists/RecentSupernovae.html} 
\hspace{2 mm}.  In 1998 the number of supernovae discovered was
160, given the naming scheme of the IAU Circulars.
In 1999 there were 203 SNe announced, but three have been shown
to be other than supernovae.}  In many cases
a low redshift SN appears at a substantial angular distance from the
galaxy or appears on a rather uniform swath of galaxy light.  In these
cases aperture photometry will give an accurate apparent magnitude for the
supernova without contamination by other sources of
photons.  One can further justify the use of aperture photometry of a
visible SN by 
obtaining an image of its host galaxy, for example, from the STSci Digital
Sky Survey ({\tt http://archive.stsci.edu/dss/dss\_form.html}) 
to gauge how problematic the location of the SN is.

In many other cases, however,  the supernova is superimposed on an
arm or the core of a galaxy, or it could be a small angular distance from
a bright foreground star in our Galaxy, so the best way to obtain the
apparent magnitude of the SN by itself is to subtract a template image of
the sky-without-supernova from the images of the sky-with-supernova.  
However, one usually does not have appropriately exposed images in all
relevant filters of such a field before the SN has appeared.  Thus, one
must wait a year or longer to obtain the galaxy template images.

Other factors contribute to the random and systematic errors of the
apparent magnitudes: the use of different filters; systematic differences
in the standard stars observed; whether the data were obtained
photographically, photoelectrically, or with CCD cameras; and whether one
reduces CCD data with aperture photometry or point spread function (PSF)
fitting.  As Suntzeff (2000) points out, even data obtained at the same
site with two instruments on two different telescopes, reduced by the same
observer using the same local standards, can give significantly different
results, owing to the non-stellar nature of SN spectra.  All of these
factors contribute to the error budget of our apparent magnitudes and our
absolute magnitudes.  Consequently, we include information in an Appendix
relating to filters and transformation equations.

\subsection {Absolute magnitudes}

In order to know the absolute magnitudes of Type Ia SNe we need to observe
a number of extinction-corrected Type Ia SNe in galaxies with known
distances. Observations of Type Ia supernovae over the past decade have
shown that while Type Ia SNe are not standard candles {\em per se}, they
are standardizable candles, forming a single-parameter family.

Phillips (1993) and Hamuy et al. (1996) noted a relationship between
the rate of decline of Type Ia SNe and their intrinsic luminosities,
characterized by the number of magnitudes these objects decline in the
first 15 days after B-band maximum ($\Delta$m$_{15}$(B)).  The Phillips
relation shows that Type Ia SNe with smaller values of
$\Delta$m$_{15}$(B) are intrinsically more luminous.  A typical value
of $\Delta$m$_{15}$(B) is 1.1 mag.

Riess, Press \& Kirshner (1996a, hereafter RPK) and Riess et al. (1998)
outlined the multi-color light curve shape (MLCS) method of fitting Type Ia
SN photometry. The key parameter, which they call $\Delta$, is basically
the number of magnitudes that a Type Ia SN is brighter than or fainter than
some fiducial object.  In other words, $\Delta \equiv$ M$_V -$
M$_V$(fiducial). Objects with $\Delta < 0$ mag are more luminous, rise and
decline more slowly after maximum light; objects with $\Delta > 0$ mag are
less luminous, rise and decline more rapidly.\footnote[5]{We make reference
in this paper to ``slow decliners'' and ``fast decliners'' and hope that no
confusion arises as a result; in fact, there is a continuum of decline
rates.} Using Cepheid calibrations of a set of nearby SNe, Jha et al.
(1999b) find M$_V$ = $-$19.34 $\pm$ 0.17 for the fiducial Type Ia SN with
$\Delta = 0.0$ mag.  The MLCS method uses the data for up to four
photometric bands (BVRI) and simultaneously gives the light curve
solutions, the reddening along the line of sight to the SN (i.e. the {\em
sum} of the reddening in our Galaxy and the reddening within the host
galaxy), the time of maximum light,\footnote[6]{Throughout this paper we
shall define the time of maximum to be the B-band maximum; V-band maximum
occurs about 2 days after B-band maximum.  Time $t$ will be the number of
days since B-band maximum.} and the distance modulus of the SN.

A third method for characterizing the light curves of Type Ia SNe is the
``stretch factor'' method of Perlmutter et al. (1997).  The stretch factor
is used to broaden or narrow the rest-frame timescale of a template light
curve.  The Phillips parameter ($\Delta$m$_{15}$(B)) and the stretch
factor ($s$) are related (Perlmutter et al.) as follows:
 
\begin {equation}
\Delta m_{15}(B) \; = \; 1.96 \pm 0.17 (s^{-1} \; - \; 1) \; \; + \; 1.07 \; .
\end {equation}

\parindent = 0 mm
 
Assumed in the stretch factor, which is used for B-band and V-band
data (but not for R and I), is the notion that slow decliners are slow
risers and vice versa.

\parindent = 9 mm 

All three empirical fitting methods are attempts to model the differences
in light curve shape, attributable to a range of absolute magnitudes,
which in turn are attributable to the details of the explosions giving
rise to the various Type Ia SNe (H\"{o}flich \& Khokhlov 1996; Nomoto,
Iwamoto \& Kishimoto 1997).

\subsection{Interstellar reddening and absorption}

If we have a sample of nearby Type Ia SNe which occurred at a significant
angular distance from any spiral arms or the central bulges of the host
galaxies, we might safely assume they were unreddened in their hosts, and
that their B$-$V colors are intrinsic to these objects. Schaefer (1995)
noted that Type Ia SNe with ``normal'' spectra and/or decline rates have
an intrinsic B$-$V = 0.00 $\pm$ 0.04 at 
maximum light.  This would imply that if a Type Ia SN has an {\em
observed} B$-$V color at maximum light of, say, 0.20, then its color
excess E(B$-$V) $\equiv$ A$_B -$ A$_V \approx$ 0.20 mag.  Using the
standard value of R$_V \equiv$ A$_V$ / E(B$-$V) = 3.1 (Sneden et al. 1978,
Rieke \& Lebofksy 1985), and assuming that dust in the host galaxy of a SN
is the same as dust in our Galaxy (Riess, Press, \& Kirshner 1996b), we
can determine the V-band extinction to the object. With an estimate of the
SN's absolute magnitude from any of the three light curve fitting methods
described above we can then determine the distance to the SN and its host
galaxy.

RPK had as one of their motivations for MLCS that there was nearly a one
magnitude range of apparent intrinsic color at maximum light for Type Ia
SNe.  Assuming that all Type Ia SNe have the same intrinsic color at
maximum actually increases the scatter of the Hubble diagram.  For
example, observations of SN 1991bg (Leibundgut et al. 1993) and our own
observations of SN 1999da (Krisciunas et al. 1999) show that these two
``fast decliners'' had B$-$V $\gtrsim$ 0.6 at the time of maximum light.  
``Slow decliners'' such as SN 1991T (Lira et al. 1998) can have excess
short wavelength (U-band and B-band) light.  The MLCS method takes into
account these differences in color to a large extent, and its prime
advantage, as already stated, is that it uses the full BVRI data set to
derive simultaneously A$_V$, the intrinsic brightness/decline rate
parameter $\Delta$, the distance, and the time of B-band maximum.

Lira (1995) and Phillips et al. (1999, in particular, their Fig. 1) have
shown that {\em all} unreddened Type Ia SNe, no matter what their decline
rates, have essentially the same B$-$V color evolution from 30 to 90 days
after V-band maximum.  (This is also shown graphically by RPK.)
Phillips et al. (1999) find

\begin {equation}
(\rm B - \rm V)_0 \; = \; 0.725 \; (\pm \; 0.05) \; - \; 0.0118 \; (t_V \;
- \; 60) \; .
\end{equation}

\parindent = 0 mm

If Eq. 3 holds for all Type Ia SNe, then we may use it and accurate
measurements of the observed B$-$V color between 30 and 90 days after
V-band maximum to determine E(B$-$V).  The primary drawback of Eq. 3 is
that by the time a typical SN is one to three months past maximum its
V-band light has dimmed by 1.5 to 3 mag and its B-band light has
diminished 2.5 to 4 mag.

\parindent = 9 mm

If we attempt to determine A$_V$ solely from B-band and V-band photometry,
any errors in the adopted color excess E(B$-$V) are scaled by R$_V
\approx$ 3.1.  More specifically, since A$_V$ = R$_V \times$ E(B$-$V), the
uncertainty in A$_V$ will be given by:

\begin {equation}
\sigma^2_{A_V} \; = \; (R_V)^2 \sigma^2_{E(B-V)} \; + \;
(E(B-V))^2 \sigma^2_{R_V} \; . 
\end{equation}

Given the infrared selective extinction coefficients determined by Rieke
\& Lebofsky (1985) we use the following relationships between color excess
and A$_V$:

\begin {equation}
A_J \; = 0.282 \; A_V \; \longrightarrow \; A_V \; = \; 1.393 \; E(V-J) \; .
\end {equation}

\begin {equation}
A_H \; = 0.175 \; A_V \; \longrightarrow \; A_V \; = \; 1.212 \; E(V-H) \; .
\end {equation}

\begin {equation}
A_K \; = 0.112 \; A_V \; \longrightarrow \; A_V \; = \; 1.126 \; E(V-K) \; .
\end {equation}

Fig. 1 shows graphically the advantage of determining A$_V$ from a
combination of optical and infrared data. Consider the {\em worst} case
scenario, that the coefficient on the left hand part of Eq. 7 is only
known to $\pm$50 percent.  Then A$_V$ = 1.126 $\pm$ 0.072 $\times$
E(V$-$K).  For the case of A$_V$ = 1.00 mag, E(V$-$K) = 0.888 (solid line
in Fig. 1). The dashed line in Fig. 1 is the {\em best} case scenario for
optical data; it assumes R$_V$ = 3.1 $\pm$ 0.1 and E(B$-$V) = 0.323 to
give A$_V$ = 1.00 mag.  We know that R$_V$ is not a universal constant
(Cardelli, Clayton, \& Mathis 1989).  Riess, Press, \& Kirshner (1996b)
find that the most appropriate value for a set of Type Ia SNe is R$_V$ =
2.55 $\pm$ 0.3.  In Fig. 1 the dotted line uses R$_V$ = 2.55 $\pm$ 0.3 and
E(B$-$V) = 0.392 to give A$_V$ = 1.00 mag. We show below that the color
excess E(V$-$K) for ``normal'' Type Ia SNe can be determined to $\pm$ 0.1
mag or better.\footnote[7]{Phillips et al. (1999) show in their Fig. 5
that except for the subluminous fast decliners, Type Ia SNe have very
similar intrinsic colors B$_{max} -$ V$_{max}$ for a wide range of decline
rates. It may not be long before Type Ia SNe are officially divided into
three sub-classes, SN 1991T-like ones, ``normal'' ones, and SN 1991bg-like
ones.  See also Branch, Fisher, \& Nugent (1993).} While one can determine
the B$-$V color of a supernova to $\pm$ 0.02 mag, determining the color
{\em excess} is another matter.

In this paper we shall show that V-band and infrared photometry (at J (1.2
$\mu$m), H (1.65 $\mu$m), and K (2.2 $\mu$m)) can give us a good handle on
A$_V$ without scaling up any errors by a factor as large as R$_V$.

\subsection{What data are available?}

Very little infrared photometry has been published on Type Ia SNe.  The
canonical papers are by Elias et al. (1981, 1985). In their second paper
(see their Fig. 7) these authors even note the small scatter ($\pm$ 0.1
mag) in V$-$K for four Type Ia SNe (1972E, 1980N, 1981B, and 1981D). Frogel
et al. (1987) present much infrared data for SN 1986G, which was a highly
reddened (E(B$-V) \geq$ 0.6 mag) ``atypical subluminous event'' (Phillips
et al. 1987, Meikle 2000).
   
Suntzeff et al. (1999) and Jha et al. (1999b) give optical data for
SN 1998bu.  Mayya, Puerari, \& Kuhn (1998) and Jha et al. (1998b)
give infrared data.  Meikle \& Hernandez (1999) show other infrared
data in graphical form, while their data values are found in Meikle
(2000) and will be discussed in full by Hernandez et al. (2000).
The photometry of SN 1998bu comprises the first large infrared
data set for a ``normal'' Type Ia SN which contains data before maximum.  
We note that the infrared data of Jha et al. (1999b) prior to B-band
maximum are systematically fainter than the infrared data of Mayya et al.
(1998) and Meikle (2000), by 0.1 mag or more.  We do not address the
issue of whose data might be systematically ``in error'' prior to
maximum light.  Perhaps the true uncertainties of their data are larger
than the errors quoted.

Here we present optical BVRI and infrared JHK data on SNe 1999aa, 1999cl,
1999cp, including infrared data on two of the three before maximum.

We shall show that the V$-$K color evolution of an ensemble of objects (SNe
1972E, 1980N, 1981B, 1981D, 1983R, 1998bu, 1999cl and 1999cp) suggests that
there is a uniform color relationship for Type Ia SNe with a range of
$\Delta$ values.  Since SNe 1972E, 1980N, 1983R, and 1999cp have suffered
minimal reddening due to our Galaxy and in their host galaxies, we can use
data for these objects and the data of supernovae that {\em are} reddened
in their hosts to produce multi-dimensional global solutions for each color
(V$-$J, V$-$H, V$-$K).  We obtain, as a desired result of the analysis, the
color excessses of each reddened object, which, with Eqs. 5 through 7,
easily yield estimates of A$_V$ $-$ necessary for an accurate
determination of the object's distance.

\section{Observations}

Supernova 1999aa in NGC 2595 was independently discovered with three
different telescopes from 11 to 13 February 1999 UT (Arbour 1999; Qiao et al.
1999; Nakano \& Kushida 1999).  Filippenko, Li, \& Leonard (1999) obtained a
spectrum on 12 February UT and noted that it was a peculiar Type Ia supernova
very similar to SN 1991T.

SN 1999cl was discovered by Papenkova et al. (1999) on 29 May 1999 UT,
roughly 17 days (according to our subsequent analysis) before maximum.  
Spectra obtained on June 4.2 UT indicated that this SN was highly reddened
by dust, with an approximate color excess of E(B$-$V) = 1.0 mag (Garnavich
et al. 1999). This SN occurred in NGC 4501 (= M 88), an SBb/Sc galaxy in
the Virgo cluster (Tully 1988).

SN 1999cp in NGC 5468 was discovered by King \& Li (1999) on 18 June UT. 
A spectrum by Jha et al. (1999a) indicated that it was a Type Ia SN well
before maximum light.

We have obtained infrared photometry of these three SNe with the Apache
Point Observatory (APO) 3.5-m telescope using GRIM II, which has a 256
$\times$ 256 pixel HgCdTe NICMOS-3 detector.  We have obtained optical
BVRI photometry with the APO 3.5-m using the CCD camera SPIcam,
which contains a backside illuminated SITe 2048 $\times$ 2048 array.
Additionally, we have obtained CCD photometry with the University of
Washington's Manastash Ridge Observatory (MRO) 0.76-m telescope using a
1024 $\times$ 1024 Ford Aerospace array.  Further information on the
filters and transformation coefficients is given in the Appendix.

We reduced all our CCD data with the {\sc iraf} data reduction package.  
This first involved subtracting the bias frames, constructing sky flats
with the median of three or more sky frames per filter per night, then
flattening and trimming the frames.  We determined the instrumental
magnitudes as aperture magnitudes, using {\bf phot} within the {\bf
apphot} package.  Calibration of the data was done within the {\bf
photcal} package, in particular with {\bf mknobsfile}, {\bf fitparams},
and {\bf evalfit}.  The resulting aperture magnitudes were used to
calculate differential magnitudes for each supernova in each filter, with
respect to local secondary standards.  These differential magnitudes were
then converted to standardized V-band magnitudes and colors using appropriate
values of the adopted magnitudes and colors of the secondary standards and
the transformation coefficients.

As mentioned in \S 1.1, unless one has images in all relevant filters of
the galaxy in question, one cannot carry out image subtraction photometry
until template images are obtained a year or more after the explosion of a
SN.  From an inspection of STSci Digital Sky Survey images of the host
galaxies of SNe 1999aa, 1999cl, and 1999cp we see no serious problems in
carrying out aperture photometry of these objects.  We agree, however,
that images of the host galaxy of SN 1999cl in particular should be
obtained in the future to determine if image subtraction significantly
changes the data.

Our infrared data were reduced with a package of scripts written by Alan
Diercks which runs in the {\sc iraf} environment.  This involved the
application of a bad pixel mask, the subtraction of median dark frames,
dividing by an appropriate sky flat for each filter, along with linearity
corrections, bias corrections, and fitting a plane to the images to correct
for any stray light.  Since the raw images were dithered short exposures,
the resulting frames had to be coadded to give adequate signal-to-noise.  
This was accomplished with Eugene Magnier's program {\sc mana} using
appropriate scripts.  Then aperture photometry was carried out with {\sc
iraf}.

The optical photometry is tied to standard stars of Landolt (1992). From data
taken on multiple photometric nights we have derived photometric sequences in
the field of each supernova (Figs. 2 through 5; Tables 1, 3 and 5). Using
differential magnitudes of the field stars with respect to each other, we found
no evidence for variability of the field stars to a level of constancy of $\pm$
0.03 mag or better per night.  Observations of the brighter field stars on the
photometric nights can then give mean values which have very small internal
errors.

Typically, we reduced the optical photometry of a SN with the magnitudes
and colors of four demonstrably constant field stars near the SN; this
usually included the brightest field stars as long as they were not
extremely red, located near the edge of the array, or near cosmetic
defects. Using secondary ``standards'' within the frame of a SN allows
frames taken under non-photometric conditions to be used. Since the
in-frame differential photometry was weighted by the reciprocals of the
squares of the errors, the calibration of the SN photometry is primarily
dependent on the brightest field stars used as local standards.  
Including several (or many) more faint field stars for the calibration
does not significantly improve the results for the SN.

Our optical photometry of SNe 1999aa, 1999cl, and 1999cp are given in
Tables 2, 4, and 6 (as nightly means).  The uncertainties, given in
parentheses, take into account the photon statistics of the raw data, the
uncertainties of the V magnitudes and colors of the photometric sequences,
and the uncertainties in the transformation coefficients to the BVRI
system.

The infrared photometry of SNe 1999aa, 1999cl, and 1999cp given in Table 7
(as nightly means) are tied to standards of Hunt et al. (1998).  Our
K-band photometry was actually taken in a K$^{\prime}$ filter, which has a
shorter effective wavelength than the regular 2.2 $\mu$m filter.
Wainscoat \& Cowie (1992), who designed this filter to give greater
signal-to-noise ratios for infrared imaging, give the transformation:
K$^{\prime} -$ K = (0.22 $\pm$ 0.03) (H$-$K).  To reduce our data we first
transformed the K-band values of the Hunt et al. standards used to
K$^{\prime}$.  We then derived K$^{\prime}$ values for our SNe.  Finally,
we transform those data using: K = 1.282 K$^{\prime} -$ 0.282 H. In the
case of SN 1999aa we did not take H-band data, but used instead the mean
H$-$K vs. age relation given by Elias et al. (1985).  At the time this
object was observed, the implied differences of K and K$^{\prime}$ were
less than 0.02 mag.  

One question naturally arises. How can we be sure that a photometric
transformation applicable to regular stars is valid for SNe?  We cannot be
absolutely sure.  The uncertainty applies to the BVRI data as well as the
near infrared photometry.  The spectra are not like a black body and
evolve rapidly, and the SN 1991T-like slow decliners have spectra which
are different than the more rapid decliners. But Type Ia SNe are
considered to be in a {\em photospheric} phase until $\approx$25 days
after B-band maximum, after which the optical light curves experience a
bend.  From +25 to +35 days the rapid decline of the light curve slows
down into an exponential decline as the object enters the nebular phase
(Jeffery et al. 1992; Bowers et al. 1997).

As in the case of the optical photometry, the infrared data were not always
taken under photometric conditions.  Sometimes it was necessary to use
observations of field stars within the SN frames or the core of the host galaxy
as secondary standards, whose magnitudes were fixed with observations of
infrared standards on photometric nights.  In the case of SN 1999aa, we
used stars 2 and 3 of the NGC 2595 photometric sequence (Table 1).
For star 2 we find J = 15.30 $\pm$ 0.03, K$^{\prime}$ = 14.94 $\pm$ 0.04,
and for star 3 we find J = 13.81 $\pm$ 0.02, K$^{\prime}$ = 13.43 $\pm$
0.03.  For SN 1999cl we used the core of M 88 as a local standard.
For a 5.7 arcsec diameter software aperture with the sky annulus of
{\em radius} 4.7 to 6.6 arcsec we find J = 11.33 $\pm$ 0.02, H = 10.52
$\pm$ 0.02, K$^{\prime}$ = 10.21 $\pm$ 0.05.  For SN 1999cp we find that
star 1 of the NGC 5468 photometric sequence (Table 5) has J = 15.18
$\pm$ 0.02, H = 14.85 $\pm$ 0.03, K$^{\prime}$ = 14.83 $\pm$ 0.08.
A second star 43 arcsec east and 18 arcsec north of star 1 has
J = 15.65 $\pm$ 0.02, H = 15.14 $\pm$ 0.02, K$^{\prime}$ = 14.83 $\pm$
0.07.

In addition to our own data for SN 1999aa, 1999cl, and 1999cp, we used
optical photometry of SN 1972E (Ardeberg \& de Groot 1983), 1980N and 1981D
(Hamuy et al. 1991), 1981B (Buta \& Turner 1983), 1983R (Tsvetkov 1988),
and 1998bu (Jha et al. 1999b). Infrared photometry of these SNe is given by
Elias et al. (1981, 1985), Mayya et al. (1998), Jha et al. (1999b), and
Meikle (2000). We only use two nights' infrared data for SN 1972E, nights
on which the data were obtained with the Palomar 200-inch telescope;
infrared data taken on other nights with a 0.6-m telescope are much less
accurate. Optical photometry for these other objects can be fit with light
curve templates derived with $\Delta$ values ranging from $-$0.38 mag (for
SN 1972E; Jha et al. 1999b) to +0.23 mag (for SN 1981D; MLCS fit to B-band
and V-band data given by Hamuy et al. 1991).

One can construct V-band templates using the MLCS V-band vectors to
determine what the likely V magnitude of a SN was at the times that any
infrared data were taken.  This is how we determined the V$-$J, V$-$H,
and V$-$K colors of SNe 1999aa, 1999cl, 1999cp and the just-mentioned
other SNe.  Because of the smoothness of V-band light curves near the
time of maximum light (see e.g.  Riess et al. 1999), in many cases this
V-band interpolation involves uncertainties conservatively estimated to
be $\pm$0.03 to $\pm$0.05 mag.

\section{Discussion}

In Table 8 we give the MLCS fits to the optical photometry of SNe 1999aa,
1999cl, and 1999cp, along with values for other objects discussed in this
paper.  Since the range of $\Delta$ for the updated MLCS method (Riess et
al. 1998) is $-$0.5 to +0.5 mag, there is no extrapolation in
$\Delta$-space involved in applying MLCS to these three objects.

SN 1999aa was an overluminous, slow decliner, with $\Delta$ = $-$0.47 $\pm$
0.08 mag. This is in accord with the pre-maximum spectrum, which was
similar to the canonical slow decliner SN 1991T (Lira et al. 1998;
Filippenko et al. 1999). From a fourth order polynomial fit to the B-band
data we find $\Delta$m$_{15}$(B) = 0.746 $\pm$ 0.024 mag.  In a recent
study of 22 Type Ia SNe Riess et al. (1999) found $\Delta$m$_{15}$(B)
values ranging from 0.86 to 1.93 mag.  Thus SN 1999aa is one of the most
slowly declining Type Ia SNe known, implying that it is one of most
intrinsically bright.  Peculiarities in its light curve at optical or
infrared wavelengths might not necessarily affect our conclusions based on
an ensemble of other objects.  MLCS fitting indicates that SN 1999aa is
unreddened in its host.  The Schlegel et al. (1998) reddening model of our
Galaxy indicates that SN 1999aa has been reddened by E(B$-$V) = 0.04 mag by
dust in our Galaxy.

MLCS fitting of the BVRI data of SN 1999cl indicates highly non-standard
reddening for this object.  Also, the late B-band and I-band data were
roughly 0.3 mag brighter than the MLCS templates.  One implication would be
that the late-time infrared data may be ``too bright'' as well. Our
photometry indicates that at the time of B-band maximum, the color of the
SN was B$-$V = 1.23 $\pm$ 0.05.  Given $\Delta$ = +0.20 mag for this
object, the implied intrinsic color at maximum light was B$-$V = +0.09.
Thus, the color excess is E(B$-$V) $\approx$ 1.14 mag.

SN 1999cp is found to be a moderately overluminous object. We find $\Delta
= -$0.31 $\pm$ 0.10 mag.  It is unreddened in its host.  The Galactic
reddening at the coordinates of this object is E(B$-V$) = 0.025 mag
(Schlegel et al. 1998).

In Figs. 6 through 8 we show the light curves for SNe 1999aa, 1999cl, and
1999cp.  It can readily be seen that the V-band templates allow us to
interpolate the V-band data such that we can obtain V $minus$
some-infrared-magnitude at the time that infrared data were taken.

In Fig. 9 we show the V {\em minus} infrared colors for SNe 1972E,
1980N, 1981B, 1981D, 1983R, 1998bu, 1999cl, and 1999cp.  The reader
will note that the data for each color index arrange themselves in
three bands. SNe 1972E, 1980N, 1981B, 1981D, 1983R, and 1999cp
constitute, at first glance, one set of relatively unreddened objects,
while SN 1998bu is one magnitude redder, and SN 1999cl is still redder.

SN 1972E has a host reddening of E(B$-$V) = 0.01 $\pm$ 0.03 mag according
to Phillips et al. (1999) $-$ in effect, zero.  SN 1983R is not well
contrained at maximum light, but at times for which Eq. 3 holds, its B$-$V
color (after correcting for Galactic reddening) indicates that it is
unreddened in its host.  MLCS indicates that SN 1999cp is unreddened in its
host.  For SN 1980N Phillips et al. (1999) give a total color excess of
E(B$-$V) = 0.07 $\pm$ 0.02 mag, of which 0.021 is due to our Galaxy.  MLCS
fits indicate that SN 1981B and 1981D have A$_V$ = 0.35 $\pm$ 0.15 and
A$_V$ = 0.44 $\pm$ 0.22 mag, respectively.  Phillips et al. (1999) give a
total color excess for SN 1981B of E(B$-$V) = 0.13 $\pm$ 0.03 mag, while
Phillips (private communication) indicates that E(B$-$V) = 0.18 $\pm$ 0.06
mag for SN 1981D.

Considering only the V$-$K and V$-$H data for the moment, the data are
most easily fit by two line segments.\footnote[8]{There is no {\em a
priori} reason to assume that the color curve could be fit by two
linear segments, but that is certainly the simplest approach to adopt.}
There exists some cross-over time when the slope of the color fit
switches from negative to positive. Let us call this T$_c$.  Then for
$t < $ T$_c$, V$-$K = a$_0$ + a$_i$ + b$_1$ ($t - $T$_c$), and for $t \geq $
T$_c$, V$-$K = a$_0$ + a$_i$ + b$_2$ ($t - $T$_c$).  Fits of the same form were
made to the V$-$H data. For V$-$J a non-linear fit (at least a second
order polynomial) is warranted for $t <$ T$_c$, so we have added a term
c($t -$ T$_c$)$^2$.  

In order to treat the data as objectively as possible, we did not
assign {\em a priori} special status to any object(s).  We used
{\sc pascal} routines given by Bevington \& Robinson (1992) and
Marquardt's method of carrying out a least-squares fit to an arbitrary
function to minimize the $\chi^2$ values of the fits, assuming initial
values of a$_0$ for each color and solving {\em simultaneously} for the
best common values of T$_c$, b$_1$, b$_2$, c and offsets (a$_i$) from 
the fiducial loci $-$ one offset for each supernova.  Note that the 
same reduced $\chi^2$ values are obtained for any other 
a$_0$.  Any shift in the value of a$_0$ for the multi-dimensional
fit for a given color shifts the resulting offsets a$_i$ by the
same amount.

In Table 9 we give our fiducial loci and have scaled the {\em
uncertainties} of the fit parameters by the square root of the reduced
$\chi^2$ values of the global fits.  We assert that by giving no special
treatment to any individual object(s) and by scaling the uncertainties
of the resultant fit parameters we make the most objective use
of the whole data set.

The reduced $\chi^2$ values of the multi-dimensional fits are 18.2,
6.23, and 3.63, respectively, for V$-$J, V$-$H, and V$-$K.  Thus, the
data and our chosen fit functions indicate a sequence of ``goodness of
fit'' as we proceed to subsequent V {\em minus} near infrared colors.
That these reduced $\chi^2$ values are greater than unity indicates
either that: (1) the photometric errors have been underestimated; (2)
there is intrinsic scatter between objects; or (3) the fit functions
were not appropriate.

In Fig. 10 we have compacted the data shown in Fig. 9, adjusting the data
of all eight SNe by the derived offsets a$_i$.  This shows the scatter of
data if all eight SNe had the mean colors of our fiducial loci. Another way
to consider the goodness of fit of the data is to determine the
root-mean-square residual of the data in Fig. 10 with respect to the
fiducial loci.  For V$-$J, V$-$H, and V$-$K, respectively, those {\sc rms}
uncertainties are $\pm$0.193, $\pm$0.105, and $\pm$0.090 mag.  Since MLCS
applied to optical data typically gives an uncertainty of $\pm$0.15 mag or
more for A$_V$, V$-$H and V$-$K photometry obtained at several epochs on
the same object have the strong potential of giving more accurate A$_V$
values.  As we have already shown in Fig. 1, for moderately to highly
reddened SNe V$-$K data will almost always give more accurate A$_V$ values
than B$-$V data.

In Table 10, column 2, we give the derived offsets a$_i$ for each of the eight
``normal'' Type Ia SNe.  We did not scale the uncertainties of the a$_i$ values
for a given color by the square root of the reduced $\chi^2$ value of the {\em
global} fit.  Rather, since different objects made different contributions per
data point to the total $\chi^2$ sum, we scaled the uncertainties of the
a$_i$'s by the square root of $\chi^2_{\nu}$ for that object, but only
if $\chi^2_{\nu}$ was greater than unity.

In column 3 of Table 10 we give the estimated Galactic reddening E(B$-$V)
from Schlegel et al. (1998).  Using the scaling factors from Eqs. 5 through
7 and a value of R$_V$ = 3.1 we give in column 4 the expected color
excesses E(V$-$J), E(V$-$H), and E(V$-$K) due to dust in our Galaxy.  
Finally, in column 5 we give the implied host reddening values (with
respect to the fiducial loci), equal to the total reddening (in column 2)
minus the Galactic reddening (in column 4).

Though any adopted values of a$_0$ would give the same reduced $\chi^2$
values for the global fits, we have adopted particular values which
establish not just fiducial loci, but {\em unreddened} loci. If we correct
the a$_i^{\prime}$ values for SN 1980N for the small amount of host
reddening (E(B$-$V) = 0.05 $\pm$ 0.02 mag) given by Phillips et al. (1999),
yielding a$_i^{\prime \prime}$ for this object, and if we let a$_i^{\prime
\prime}$ = a$_i^{\prime}$ for SNe 1972E, 1983R, and 1999cp, then the
weighted means of the a$_i^{\prime \prime}$ values are all 0.00 $\pm$ 0.03
based on these four objects, color by color.  

Concerning SNe 1981B and 1981D, there is no evidence from VJHK analysis that
these objects are reddened.  This contradicts the implications of fits to
BV(R) data mentioned above which indicate that they have non-zero reddening.

SNe 1980N and 1981D occurred in the same galaxy, thus they must have the
same distance modulus.  SN 1980N had an observed peak magnitude of
V$_{max}$ = 12.43 $\pm$ 0.03, while SN 1981D had V$_{max}$ = 12.41 $\pm$
0.03.  Their $\Delta$ values (see Table 8) were almost identically the
same, implying that they had the same absolute magnitudes.  Yet SN 1981D
has been reddened in B$-$V by 0.11 $\pm$ 0.06 mag more than SN 1980N
according to the analysis of Phillips et al. (1999) and Phillips (private
communication). We must conclude either that there is some hidden
systematic error in the B-band photometry of SN 1981D or that something
other than reddening by dust can affect the color of a Type Ia SN (Tripp \&
Branch 1999).

Had we established the unreddened loci with the values of a$_i^{\prime}$ for
SNe 1972E, 1983R, 1999cp, 1981B, and 1981D, the zero points a$_0$ would have
to be changed less than 0.03 mag in order for the weighted means of
a$_i^{\prime}$ to be zero for V$-$H and V$-$K. The V$-$J colors are more
problematic, and we believe the data may be showing evidence of scatter
between different objects.  Given the unresolved mystery of the implied host
galaxy extinctions of SNe 1981B and 1981D, we feel it is better for now to
fix the zero points of the unreddened loci via SNe 1972E, 1980N, 1983R, and
1999cp and to take new data on new objects.

In the case of SN 1998bu our derived color excesses and Eqs. 5 through 7
give A$_V$ = 1.038 $\pm$ 0.166, 1.165 $\pm$ 0.084, and 1.137 $\pm$ 0.055
mag from the V$-$J, V$-$H, and V$-$K data, respectively.\footnote[9]{Here,
and in the discussion that follows, we assume a $\pm$20 percent uncertainty
in the ratio of A$_{\lambda}$/A$_V$ for the infrared bands.} These are
estimates of the total V-band extinction, of which roughly 0.074 mag is due
to dust in our Galaxy. The weighted mean of our three estimates is A$_V$ =
1.138 mag, with an uncertainty conservatively estimated to be equal to the
smallest of the errors of the three estimates, or $\pm$ 0.055 mag.  Jha et
al. (1999b) obtain A$_V$ = 0.94 $\pm$ 0.15 from MLCS fits of BVRI data;
these authors also compared optical and infrared data of SN 1998bu to SNe
1980N, 1981D, and 1989B, much as we have done here, obtaining A$_V$ = 0.9
$\pm$ 0.2.  Suntzeff et al. (1999) adopt a Galactic reddening of E(B$-$V) =
0.025 $\pm$ 0.003 mag and obtain a host galaxy reddening of E(B$-$V) = 0.34
$\pm$ 0.03 mag.  Assuming R$_V \equiv$ 3.1 implies A$_V$ = 1.132 $\pm$
0.093 mag. Meikle (2000) adopts E(B$-$V) = 0.35 $\pm$0.03 mag for the total
color excess, implying A$_V$ = 1.085 $\pm$ 0.093 mag. Our value of A$_V$
for this supernova and an adopted color excess E(B$-$V) = 0.358 $\pm$ 0.030
mag gives R$_V$ = 3.18 $\pm$ 0.31. Our VJHK analysis is in good agreement
with the results based on BV photometry and a Galactic value of R$_V$.

For SN 1999cl our derived color excesses give estimates of the total
extinction A$_V$ = 2.003 $\pm$ 0.407, 1.955 $\pm$ 0.164, and 2.036 $\pm$
0.108 mag from the V$-$J, V$-$H, and V$-$K data, respectively.  The
weighted mean is A$_V$ = 2.011 $\pm$ 0.108 mag (taking the most
conservative estimate of the uncertainty).  Since roughly 0.118 mag is due
to our Galaxy, the implication is that the V-band extinction due to dust in
the host galaxy is 1.89 mag.

As mentioned above, for SN 1999cl B$-$V = 1.23 $\pm$ 0.05 at $t$ = 0, and
E(B$-$V) $\approx$ 1.14 mag. If R$_V$ = 3.1 applies to this object, then
A$_V$ = 3.53 mag, significantly different than our result of A$_V$ = 2.01
$\pm$ 0.11. A$_V$ = 3.53 leads to a distance estimate for M 88 of 7.2 Mpc,
clearly incorrect if it is in the Virgo cluster (see below).  A$_V$ = 2.01
mag and R$_V$ = 3.1 would imply that E(B$-$V) = 0.65 mag and that the
intrinsic color of this SN at maximum light was B$-$V = 0.58.  However,
that contradicts the implied intrinsic color at maximum light of B$-$V =
+0.09 for a SN with $\Delta$ = +0.20 mag. Another alternative would be that
R$_V$ = 2.01 / 1.14 $\approx$ 1.8, which strikes us as unrealistic if the
composition of the dust in M 88 is anything like that of dust in our
Galaxy.  Cardelli et al. (1989) indicate that R$_V$ can be as low as 2.60
in our Galaxy, while Riess et al. (1996b) give R$_V$ = 2.55 $\pm$ 0.3 as
the best reddening ratio applicable to a set of Type Ia SNe.  R$_V$
$\approx$ 1.8 would mean very small dust grains that would highly redden
the light. We conclude that M 88 either has very unusual dust, or that
there is a second contributor to the observed color of SN 1999cl in
addition to reddening by dust (Tripp \& Branch 1999).

From the MLCS V-band fit for SN 1999cl we adopt V$_{max}$ = 13.68 $\pm$
0.08. $\Delta$ = +0.20 mag implies that M$_V \approx -$19.14, for which we
adopt an uncertainty of $\pm$ 0.3 mag.  With A$_V$ = 2.01 $\pm$ 0.11 mag,
the implied distance modulus is m$-$M = 30.81 $\pm$ 0.33 mag.  The distance
to SN 1999cl and its host, M 88, is 14.5$^{+2.4}_{-2.0}$ Mpc.

For comparison let us derive a distance based on the infrared data and the
calibration of absolute infrared magnitudes of SNe given by Meikle (2000),
At $t$ = 13.75 days he gives M$_J$ = $-$16.86 $\pm$ 0.06, M$_H$ = $-$18.22
$\pm$ 0.05, and M$_K$ = $-$18.23 $\pm$ 0.05 for six ``normal'' Type Ia SNe
(i.e. he has excluded the fast decliner SN 1986G and the slow decliner SN
1991T).  We shall assume that SN 1999cl has infrared absolute magnitudes
equal to the just-mentioned mean values, but with uncertainties of
$\pm$0.15 mag. From an interpolation of our near-infrared photometry we
estimate that SN 1999cl had J = 14.07 $\pm$ 0.05, H = 12.93 $\pm$ 0.05, K =
12.79 $\pm$ 0.05 at $t$ = 13.75 days.  Assuming A$_V$ = 2.01 $\pm$ 0.11 and
the coefficients in Eqs. 5 through 7, the corresponding extinctions would
be A$_J$ = 0.57, A$_H$ = 0.35, and A$_K$ = 0.23 mag, with uncertainties of
$\pm$ 0.02 or more.  The resulting estimates of the distance modulus are
m$-$M = 30.36, 30.80, and 30.79 mag (from J, H, and K), with uncertainties
of $\pm$0.16 mag. Thus, the H and K data give a distance modulus of 30.8
mag, which can be compared to the value from MLCS (30.83 $\pm$ 0.6:) and
the value from VJHK analysis (30.81 $\pm$ 0.33).  The {\em most likely}
values of the distance modulus, determined by means of different methods,
agree very well.  The prime uncertainty in the distance comes from the
uncertainty in the absolute magnitude of SN 1999cl (be it M$_V$ or in the
infrared).

Our estimate of the distance to M 88 agrees within the errors with modern
estimates of the distance to the Virgo cluster (Pierce, McClure, \& Racine
1992; Ferrarese et al. 1996; Gibson et al. 2000).  Given that M 88 is 2.0
degrees from M 87 (which can be considered the core of the Virgo cluster),
the projected distance of M 88 from M 87 is 0.5 Mpc.  Thus, the derived
distance and the angular proximity of M 88 to M 87 is consistent with M 88
being in the Virgo cluster.

What of overluminous, slowly declining SNe and underluminous, rapidly
declining ones? In Fig. 11 we show the V$-$J and V$-$K colors of SNe
1999aa, and also the V$-$K colors of SN 1986G.  Clearly, the data do {\em
not} follow the photometry of ``normal'' Type Ia SNe in the mid-range of
decline rates.  Thus, we predict there will be different V
{\em minus} near infrared color loci for Type Ia SNe of different
intrinsic luminosities.

We note again the work of Tripp \& Branch (1999, and references therein)
on the question of a ``second parameter'' (i.e. a descriptor for Type Ia
SNe in addition to the decline rate).  SN models (H\"{o}flich \& Khokhlov
1996; Nomoto et al. 1997) suggest that the luminous slow decliners may be
double degenerate explosions; putting two logs in the fireplace can give a
brighter fire that lasts longer.  The second parameter may be related to
the contribution of a particular species (such as cobalt, iron, or sulfur)
to the opacity of the explosion (see Bowers et al. 1997). A second
parameter might explain the very low values of R$_V$ obtained for SNe like
1999cl.  It might also explain deviations from our color curves such as
the late-type behavior of SN 1999cl and the early time behavior of SN
1991T, which exhibited a K-band ``excess'' of $\approx$1.5 mag at $t$ =
$-$11.3 days (see Meikle 2000).

The existence of well behaved V-band light curves (e.g. from MLCS) and V
{\em minus} infrared color relations implies the existence of infrared
light curve templates. In Fig. 12 we show the predicted K-band light curve
templates (good to $\pm$ 0.2 mag) for the range of $\Delta$ implied by the
unreddened locus in Fig. 10. In Fig. 13 we show the J-, H-, and K-band
light curve templates for the fiducial Type Ia SN with $\Delta$ = 0 mag.  
These infrared light curve templates are also listed in Table 11 and may be
considered an improvement on those based solely on the data given by Elias
et al. (1981, 1985). Figs. 12 and 13 also confirm what Meikle (2000) showed
convincingly for the first time, that the near infrared flux peaks several
days before B-band maximum.

\section{Conclusions}

Using new data and data from the literature, we have outlined a method for
determining the V-band extinction towards Type Ia SNe. For SNe with $-0.4
\lesssim \Delta \lesssim +0.2$ mag there appears to be uniform evolution
of V {\em minus} infrared colors from (at least) one week before B-band
maximum until four weeks after B-band maximum. The slow decliner SN 1999aa
(with $\Delta$ = $-$0.47 mag) did not exhibit the same color evolution as
eight other Type Ia SNe with faster decline rates.

If the V {\em minus} infrared color relations discussed here are
``universal'', the implication is that one can determine A$_V$ for a {\em
normal} Type Ia SN using an interpolated V-band light curve and a small
amount of infrared photometry (H or K, but maybe not J) obtained between
$-9 \leq t \leq +27$ days. Analogous statements cannot be made for the
very slow decliners like SN 1991T and the very fast decliners like SN
1991bg.

Our VJHK analysis of the moderately reddened SN 1998bu gives a value
of A$_V$ = 1.14 $\pm$ 0.06 mag, in excellent agreement with (and a possible
improvement upon) the results from optical photometry alone.  

We were led to consider this method for determining A$_V$ owing to the
challenging case of SN 1999cl.  The VJHK data available lead to a value of
A$_V$ = 2.01 $\pm$ 0.11 mag.  Along with the V-band maximum and the implied
absolute magnitude from the MLCS fits, we obtain a distance to this object
of 14.5$^{+2.4}_{-2.0}$ Mpc, in reasonable agreement with modern estimates
of the distance to the Virgo cluster.

We have several recommendations for future work: (1) We should obtain
images of M 88 after SN 1999cl has sufficiently faded away so that image
subtraction techniques can be used for this object.  (2) Fast decliners
(such as SN 1991bg and 1999da, which occurred in elliptical galaxies)
should be observed in the infrared as well as the optical. (3) More
unreddened slow decliners should be observed from one week (or more) before
until four weeks after maximum light to see how their behavior differs from
the more normal Type Ia SNe.  Given that there are families of BVRI light
curves, it would not be surprising if there are families of V {\em minus}
infrared color curves.  (4) The next generation of MLCS and other light
curve fitting algorithms should use optical {\em and} infrared data.

\vspace {1 cm}

\acknowledgments

This paper is based on observations obtained with the Apache
Point Observatory 3.5-meter telescope, which is owned and operated
by the Astrophysical Research Consortium.  {\sc iraf} is a product
of the National Optical Astronomy Observatories, which is operated
by the Association of Universities for Research in Astronomy, Inc.,
under cooperative agreement with the National Science Foundation.

We thank Arne Henden (USNO) for photometric calibrations of the NGC 5468
field for SN 1999cp, which augmented our own calibration.  Eugene Magnier
and Alan Diercks developed new {\sc iraf} scripts and other software
for infrared data reduction; they took a majority of the raw infrared data
for SNe 1999aa, 1999cl and 1999cp and also provided many comments on the
original draft of this paper. Some of the raw CCD data for SNe 1999cl and
1999cp were obtained by John Armstrong, Guillermo Gonzalez, and Chris
Laws. Brian Skiff kindly determined the coordinates of the stars of the
photometric sequences.  We thank Craig Hogan, Bruce Margon, Brent Tully,
Bradley Schaefer, Mark Phillips, and Bruno Leibundgut for useful
discussions. We thank Peter Meikle for very useful discussions and for
making more 1998bu data available prior to publication.  We thank an
anonymous referee for many useful suggestions.  Finally, we are grateful
to the Packard Foundation and the National Science Foundation (through
grant AST-9512594) for their support.

\newpage 

\begin {center}
{\bf Appendix}

\vspace {3 mm}

{\bf Filters and Transformation Coefficients}
\end {center}

The SPIcam broad band filters used at APO are described (with graphical
and numerical transmission curves) at:

\parindent = 0 mm

{\tt http://www.apo.nmsu.edu/Instruments/filters/broad.html}

The GRIM II manual (Watson, Hereld, \& Rauscher 1997) can be obtained at:

{\tt http://www.apo.nmsu.edu/Instruments/GRIM2/}

It shows transmission curves for the APO near infrared filters.

\parindent = 9 mm

The MRO filter transmission curves are not readily available.  The B, V,
and R filters were made by Fish-Schurman, New Rochelle, New York.  Each
filter consists of pieces of Schott glass cemented together (but the cement
used is unknown).  The filter prescriptions are as follows. B: 1 mm thick
BG-12 glass, 2 mm BG-39, 1 mm GG-385; V:  2 mm GG-495, 2 mm BG-39; R: 2 mm
OG-570, 2 mm KG-3.  The I filter was obtained from Barr Associates, Inc.,
Westford, Massachusetts.  It is an interference filter with the following
specifications (matching those used by NOAO): central wavelength = 8290
$\pm$ 100 \AA; bandwith = 1950 $\pm$ 100 \AA; flatness = 5 waves;
parallelism less than 2 arcmin; blocking outside of bandpass = 10$^{-4}$
from 3000 to 11000 \AA.

In Table 12 we give the mean transformation coefficients used for
reducing the APO and MRO data.  Here $\epsilon _v$ is the scale factor
which is used to multiply the instrumental B$-$V colors to obtain
standardized V-band magnitudes.  The $\mu _{ij}$ coefficients are the
scale factors used to transform the instrumental colors to standardized
colors. Because the APO coefficients are derived from different Landolt
(1992) standards at different times of the year, it is not possible to
separate any seasonal effect from systematic differences in the zero
points of the standards.  The SN 1999aa data, all taken at APO, were
reduced with coefficients obtained from data of 1 March through 7 April
1999.  The SN 1999cl and 1999cp data taken at APO were reduced with
coefficients derived from data of 15 May through 10 June which are not
very much different.  The MRO coefficients were derived from
observations of 4 June through 9 July 1999.  Note in Fig. 7 how well the
early time B and V data of SN 1999cl fit together, though they were
obtained with two telescopes and significantly different values of 
$\mu_{bv}$.

%\end{document}

%\end{document}

\newpage 
\begin{deluxetable}{ccccccc}
\tablewidth{0pc}
\tablecaption{NGC 2595 Photometric Sequence for SN 1999aa}
\tablehead{
\colhead{Star} & \colhead{$\alpha$ (2000)} &
\colhead{$\delta$ (2000)} & \colhead {V} & \colhead {B$-$V} &
\colhead{V$-$R} & \colhead{V$-$I} }
\startdata
  SN &  8:27:42.0 & +21:29:15 &                &                &                &                \nl
   2 &  8:27:43.3 & +21:29:52 & 16.762 (0.010) &  0.849 (0.015) &  0.434 (0.011) &  0.847 (0.012) \nl
   3 &  8:27:38.2 & +21:29:55 & 15.079 (0.009) &  0.683 (0.010) &  0.373 (0.008) &  0.752 (0.008) \nl
   4 &  8:27:43.4 & +21:30:34 & 20.039 (0.026) &  1.620 (0.108) &  1.170 (0.038) &  2.626 (0.023) \nl
\nl
   5 &  8:27:40.7 & +21:31:08 & 18.979 (0.010) &  0.753 (0.018) &  0.423 (0.010) &  0.833 (0.014) \nl
   6 &  8:27:37.3 & +21:30:58 & 16.624 (0.011) &  1.070 (0.012) &  0.592 (0.012) &  1.130 (0.011) \nl
   7 &  8:27:35.2 & +21:29:47 & 18.484 (0.014) &  1.403 (0.021) &  0.809 (0.015) &  1.594 (0.014) \nl
\nl
   9 &  8:27:32.5 & +21:28:34 & 18.425 (0.014) &  0.529 (0.019) &  0.298 (0.010) &  0.676 (0.016) \nl
  10 &  8:27:35.7 & +21:28:00 & 18.516 (0.013) &  1.533 (0.022) &  0.984 (0.010) &  2.141 (0.013) \nl
  11 &  8:27:36.9 & +21:27:31 & 18.737 (0.018) &  0.937 (0.019) &  0.493 (0.021) &  1.003 (0.015) \nl
\nl
  12 &  8:27:40.8 & +21:27:22 & 15.792 (0.009) &  0.844 (0.011) &  0.453 (0.008) &  0.904 (0.009) \nl

\enddata
\end{deluxetable}

\begin{deluxetable}{ccccc}
\tablewidth{0pc}
\tablecaption{BVRI Photometry of SN 1999aa$^a$}
\tablehead{
\colhead{JD $-$ 2,451,000} & \colhead{V} &
\colhead{B$-$V} & \colhead{V$-$R} &
\colhead{V$-$I} }
\startdata
  228.5901 &  14.980  (0.004) & $-$0.014  (0.005) & $-$0.057  (0.004) & $-$0.261  (0.004) \nl
  234.6081 &  14.857  (0.004) &    0.077  (0.006) & $-$0.042  (0.006) & $-$0.442  (0.005) \nl
  238.5953 &  14.925  (0.005) &    0.119  (0.007) & $-$0.062  (0.005) & $-$0.545  (0.006) \nl
  240.6058 &  14.978  (0.004) &    0.149  (0.006) & $-$0.072  (0.007) & $-$0.542  (0.006) \nl
  244.6252 &  15.189  (0.004) &    0.211  (0.004) & $-$0.133  (0.004) & $-$0.574  (0.004) \nl
\nl
  249.5977 &  15.502  (0.004) &    0.420  (0.004) & $-$0.115  (0.004) & $-$0.383  (0.004) \nl
  252.6033 &  15.650  (0.004) &    0.620  (0.004) & $-$0.020  (0.005) & $-$0.180  (0.004) \nl
  257.6257 &  15.878  (0.003) &    0.912  (0.004) &                   &    0.156  (0.004) \nl
  263.6145 &  16.151  (0.006) &    1.184  (0.014) &    0.312  (0.008) &    0.472  (0.013) \nl
  265.6119 &  16.269  (0.005) &    1.150  (0.010) &    0.343  (0.005) &    0.571  (0.006) \nl
\nl
  274.7846 &  16.767  (0.011) &    1.044  (0.086) &                   &    0.533  (0.013) \nl
  275.6135 &  16.796  (0.004) &    1.057  (0.006) &    0.288  (0.006) &    0.525  (0.004) \nl
  279.6524 &  16.934  (0.004) &    0.997  (0.005) &    0.265  (0.004) &    0.463  (0.005) \nl
  318.6274 &  17.900  (0.014) &  0.530  (0.020)  & 0.004  (0.021) & $-$0.187 (0.037) \nl
  322.6412 &  18.050  (0.044) &  0.640  (0.053)  & 0.052  (0.067) & $-$0.264 (0.060) \nl
 \nl
\enddata
\tablenotetext{a} {The data were derived with respect to stars 2, 3, 6, and 12 of
the NGC 2595 photometric sequence.}
\end{deluxetable}

\begin{deluxetable}{ccccccc}
\tablewidth{0pc}
\tablecaption{NGC 4501 Photometric Sequence for SN 1999cl}
\tablehead{
\colhead{Star} & \colhead{$\alpha$ (2000)} &
\colhead{$\delta$ (2000)} & \colhead {V} & \colhead {B$-$V} &
\colhead{V$-$R} & \colhead{V$-$I} }
\startdata

SN & 12:31:56.0 & +14:25:35 &                 &                 &                 &                 \nl
 1 & 12:31:57.2 & +14:26:15 & 17.372  (0.005) &  0.096  (0.003) &  0.107  (0.019) &  0.163  (0.015) \nl
 2 & 12:31:55.5 & +14:26:13 & 18.450  (0.006) &  0.456  (0.007) &  0.302  (0.012) &  0.616  (0.021) \nl
 4 & 12:31:52.8 & +14:23:52 & 19.234  (0.006) &  0.491  (0.012) &  0.323  (0.052) &  0.710  (0.011) \nl
 5 & 12:31:49.1 & +14:23:38 & 18.584  (0.005) &  1.460  (0.021) &  0.929  (0.004) &  1.838  (0.009) \nl
\nl
 7 & 12:31:57.1 & +14:28:52 & 11.879  (0.004) &  0.783  (0.020) &  0.448  (0.028) &  0.817  (0.050) \nl
 8 & 12:32:08.8 & +14:28:44 & 14.687  (0.020) &  0.581  (0.030) &  0.405  (0.009) &  0.729  (0.011) \nl
 9 & 12:31:49.4 & +14:21:48 & 16.293  (0.017) &  0.900  (0.078) &  0.547  (0.030) &  0.986  (0.032) \nl
11 & 12:32:05.2 & +14:23:34 & 13.628  (0.004) &  0.841  (0.008) &  0.538  (0.005) &  0.988  (0.005) \nl
12$^a$ & 12:32:04.7 & +14:23:16 & 14.970  (0.008) &  0.716  (0.022) &  0.467  (0.015) & 0.864  (0.017) \nl
\nl
\enddata
\tablenotetext{a}{Number 12 is actually a close {\em pair} of stars.}
\end{deluxetable}

\begin{deluxetable}{cccccc}
\tablewidth{0pc}
\tablecaption{BVRI Photometry of SN 1999cl$^a$}
\tablehead{
\colhead{JD $-$ 2,451,000} & \colhead{Observatory} &
\colhead{V} & \colhead{B$-$V} & \colhead{V$-$R} &
\colhead{V$-$I} }
\startdata
  336.6966 &  APO & 14.215  (0.007) &  1.069  (0.011) &  0.630  (0.012) &  0.996  (0.014) \nl
  337.7858 &  APO & 14.138  (0.014) &  1.072  (0.020) &  0.560  (0.023) &  0.873  (0.025) \nl
  339.7589 &  MRO & 13.932  (0.008) &  1.157  (0.023) &  0.600  (0.016) &  0.919  (0.019) \nl
  340.6926 &  APO & 13.902  (0.006) &  1.127  (0.009) &  0.511  (0.009) &  0.772  (0.010) \nl 
  363.7404 &  MRO & 14.604  (0.010) &  2.184  (0.071) &  0.780  (0.015) &  1.455  (0.016) \nl
  365.7448 &  MRO & 14.727  (0.008) &  2.027  (0.037) &  0.828  (0.015) &  1.566  (0.017) \nl
  367.7305 &  MRO & 14.768  (0.018) &  2.033  (0.177) &  0.784  (0.025) &  1.647  (0.024) \nl
  368.7221 &  MRO & 14.936  (0.023) &  2.029  (0.105) &  0.934  (0.033) &  1.717  (0.032) \nl
 \nl
\enddata
\tablenotetext{a} {The APO data were reduced with respect to stars 1, 2, and 5
of the NGC 4501 photometric sequence, while the MRO data were reduced with respect
to stars 7, 8, 11, and 12.}
\end{deluxetable}

\clearpage

\begin{deluxetable}{ccccccc}
\tablewidth{0pc}
\tablecaption{NGC 5468 Photometric Sequence for SN 1999cp}
\tablehead{
\colhead{Star} & \colhead{$\alpha$ (2000)} &
\colhead{$\delta$ (2000)} & \colhead {V} & \colhead {B$-$V} &
\colhead{V$-$R} & \colhead{V$-$I} }
\startdata
SN & 14:06:31.3 & $-$5:26:49  &                 &                 &                 &                 \nl
 1 & 14:06:28.6 & $-$5:27:33  & 16.356  (0.005) &  0.525  (0.016) &  0.356  (0.004) &  0.670  (0.005) \nl
 2 & 14:06:29.4 & $-$5:28:54  & 15.124  (0.018) &  0.133  (0.006) &  0.132  (0.004) &  0.271  (0.004) \nl
 3 & 14:06:27.3 & $-$5:30:23  & 14.474  (0.010) &  0.534  (0.010) &  0.355  (0.007) &  0.720  (0.008) \nl
 4 & 14:06:32.3 & $-$5:30:45  & 15.255  (0.010) &  0.626  (0.010) &  0.399  (0.012) &                 \nl
 5 & 14:06:34.3 & $-$5:29:33  & 16.715  (0.020) &  0.610  (0.020) &  0.421  (0.047) &  0.856  (0.055) \nl
 6 & 14:06:41.6 & $-$5:26:00  & 14.617  (0.010) &  0.961  (0.010) &  0.577  (0.007) &  1.075  (0.008) \nl
 7 & 14:06:28.2 & $-$5:24:51  & 16.739  (0.043) &  1.044  (0.044) &  0.606  (0.006) &  1.136  (0.006) \nl
 8 & 14:06:23.4 & $-$5:25:46  & 15.340  (0.004) &  0.706  (0.008) &  0.432  (0.004) &  0.792  (0.004) \nl
\nl
\enddata
\end{deluxetable}

\begin{deluxetable}{cccccc}
\tablewidth{0pc}
\tablecaption{BVRI Photometry of SN 1999cp$^a$}
\tablehead{
\colhead{JD $-$ 2,451,000} & \colhead{Observatory} &
\colhead{V} & \colhead{B$-$V} & \colhead{V$-$R} &
\colhead{V$-$I} }
\startdata
  355.7334 & MRO & 14.671  (0.016)  & $-$0.116  (0.044) &    0.127  (0.020) &    0.088  (0.023) \nl
  365.7721 & MRO & 14.045  (0.005)  &    0.006  (0.008) & $-$0.023  (0.007) & $-$0.401  (0.013) \nl
  367.7459 & MRO & 14.099  (0.011)  &    0.100  (0.024) & $-$0.034  (0.014) & $-$0.412  (0.018) \nl
  368.7429 & MRO & 14.125  (0.008)  &    0.078  (0.015) & $-$0.024  (0.009) & $-$0.494  (0.015) \nl
  383.6373 & APO & 14.939  (0.004)  &    0.621  (0.006) &    0.003  (0.003) & $-$0.136  (0.004) \nl 
 \nl
\enddata
\tablenotetext{a} {The MRO data were reduced with respect to stars 2, 3, 6, and 8 of the
NGC 5468 photometric sequence, while the APO data were reduced with respect to stars
2 and 8.}
\end{deluxetable}

\begin{deluxetable}{ccccc}
\tablewidth{0pc}
\tablecaption{Infrared Photometry}
\tablehead{
\colhead{Object} &
\colhead{JD $-$ 2,451,000} & \colhead{J} &
\colhead{H} & \colhead{K} }
\startdata
SN 1999aa & 239.82 & 16.28 (0.04) &              & 15.78 (0.07) \nl 
          & 244.82$^a$ & 17.10 (0.08) &              &              \nl
          & 252.72$^a$ & 17.00 (0.06) &              & 15.90 (0.05) \nl
          & 259.59$^a$ & 16.87 (0.04) &              & 15.96 (0.05) \nl
 \nl
SN 1999cl & 337.67 & 12.96 (0.02) &              & 12.74 (0.02) \nl
          & 340.65 & 12.80 (0.02) & 12.98 (0.02) & 12.58 (0.02) \nl
          & 341.68 & 12.89 (0.02) & 13.02 (0.02) & 12.59 (0.02) \nl
          & 351.78$^b$ & 14.27 (0.04) & 13.29 (0.04) & 13.11 (0.06) \nl
          & 360.63 & 14.01 (0.02) & 12.84 (0.02) & 12.70 (0.03) \nl
          & 367.62$^b$ & 13.49 (0.04) & 12.77 (0.04) & 12.77 (0.05) \nl
\nl
SN 1999cp & 354.81 & 14.88 (0.03) & 15.04 (0.07) & 14.94 (0.15) \nl
          & 360.67$^c$ & 14.50 (0.02) & 14.77 (0.02) & 14.57 (0.06) \nl
\nl
\enddata
\tablenotetext{a} {Reduced with respect to stars 2 and 3 of the NGC 2595 
sequence. $^b$Photometry calibrated using the core of NGC 4501.
$^c$Reduced with respect to star 1 of the NGC 5468 photometric
sequence, and with respect to a second star 43 arcsec east and 18
arcsec north of star 1.  See text for details.}
\end{deluxetable}

\begin{deluxetable}{ccccccc}
\tablewidth{0pc}
\tablecaption{Light Curve Fitting Parameters}
\tablehead{
\colhead{Object} & \colhead{T(B$_{max}$)$^a$} &
\colhead {$\Delta$} &
\colhead{M$_V$} & \colhead{A$_V$} & \colhead{m$-$M} &
\colhead{Refs$^b$} }
\startdata
SN 1972E   & 41447.5(1.0) & $-$0.38(0.13) & $-$19.80(0.29) & 0.15(0.15) & 28.08(0.26) & 1 \nl
SN 1980N   & 44586.5(0.5) &   +0.20(0.20) & $-$19.14(0.20) & 0.22(0.06) & 31.44(0.32) & 2,3,4 \nl
SN 1981B   & 44672.0(0.2) & $-$0.34(0.18) & $-$19.46(0.23) & 0.35(0.15) & 31.10(0.13) & 1 \nl
SN 1981D   & 44680.5(0.5) &   +0.23(0.20) & $-$19.11(0.20) & 0.44(0.22) & 31.03(0.30) & 4 \nl
SN 1983R$^c$   & 45607.0(3.0) & $-$0.20       &            & 0.23       &             & 4 \nl
SN 1998bu  & 50952.8(0.8) &   +0.02(0.18) & $-$19.42(0.22) & 0.94(0.15) & 30.37(0.12) & 1 \nl
SN 1999aa  & 51232.37(0.23) & $-$0.47(0.08) & $-$19.80(0.15) & 0.00(0.15) & 34.53(0.16) & 4 \nl
SN 1999cl$^d$  & 51345.02(0.20) &   +0.20(0.3:) & $-$19.14(0.3:) & 2.00(0.5:) & 30.83(0.6:) & 4 \nl
SN 1999cp  & 51363.62(0.28) & $-$0.31(0.10) & $-$19.65(0.10) & 0.00(0.12) & 33.56(0.13) & 4 \nl
 \nl
\enddata
\tablenotetext{a} {Julian Date {\em minus} 2,400,000.
$^b$References: 1 (Jha et al. 1999b), 2 (Hamuy et al. 1991), 
3 (Phillips et al. 1999), 4 (this paper). 
$^c$For SN 1983R we give only the $\Delta$ value used to
make a template for interpolating the scanty V-band photometry.  Schlegel
et al. (1998) give Galactic reddening at its coordinates of
E(B$-$V) $\approx$ 0.074 mag, yielding a minimum A$_V$ = 0.23 mag.
The data and Eq. 3 indicate that SN 1983R is unreddened in its host.
$^d$See text for more accurate values of extinction and distance
for SN 1999cl.}
\end{deluxetable}

\begin{deluxetable}{lcccccc}
\tablewidth{0pc}
\tablecaption{(V $-$ [Near Infrared]) Fiducial Loci$^a$}
\tablehead{
\colhead{Color} & \colhead{$a_0$} & 
\colhead{T$_c$} & \colhead{$b_1$} & \colhead{$c$} &
\colhead{$b_2$} & \colhead{$\chi^2_{\nu}$}}
\startdata
V$-$J & $-$1.879 & 9.52(72) & $-$0.1733(0328) & $-$0.00420(00179) & 0.0851(0123) & 18.2 \nl 
V$-$H & $-$1.189 & 5.19(46) & $-$0.0613(0083) &  0.0 & 0.0840(0042) & 6.23 \nl
V$-$K & $-$0.986 & 6.44(50) & $-$0.0465(0056) &  0.0 & 0.0683(0036) & 3.63 \nl
\nl
\enddata   
\tablenotetext{a}{Time $t$ is the number of days since B-band maximum.
The least-squares fits are of the form: V $-$ [J, H, or K] = 
$a_0$ + a$_i$ + $b_1 (t - $T$_c)$ + $c (t - $T$_c)^2$ for $t <$ T$_c$ and
V $-$ [J, H, or K] = $a_0$ + a$_i$ + $b_2 (t - $T$_c)$ for $t \geq$ T$_c$,
where each supernova has its own a$_i$ value. 
The raw uncertainties from the multi-dimensional fits have been
scaled by the square root of the reduced $\chi^2$ values of the last
column.}
\end{deluxetable}

\begin{deluxetable}{lcccc}
\tablewidth{0pc}
\tablecaption{Derived Color Excesses$^a$}
\tablehead{
\colhead{Object} & \colhead{a$_i$} & \colhead{E(B$-$V)$^b$} &
\colhead{E(V$-$[J, H, or K])} &
\colhead{a$_i^{\prime}$} }
\startdata

V$-$J:    &            &                &            &               \nl
SN 1972E & $-$0.186(0.118) & 0.056(0.006) & 0.125(0.012) & $-$0.311(0.118) \nl
SN 1980N & 0.208(0.032) & 0.021(0.002) & 0.047(0.005) & 0.161(0.032) \nl
SN 1983R & $-$0.065(0.125) & 0.074(0.007) & 0.165(0.016) & $-$0.230(0.126) \nl
SN 1999cp & 0.091(0.043) & 0.025(0.003) & 0.056(0.006) & 0.035(0.043) \nl
\nl
SN 1981B & $-$0.235(0.113) & 0.018(0.002) & 0.040(0.004) & $-$0.275(0.113) \nl
SN 1981D & $-$0.155(0.053) & 0.021(0.002) & 0.047(0.005) & $-$0.202(0.053) \nl
SN 1998bu & 0.745(0.103) & 0.024(0.002) & 0.053(0.005) & 0.692(0.103) \nl
SN 1999cl & 1.438(0.269) & 0.038(0.004) & 0.085(0.008) & 1.353(0.269) \nl
\nl

V$-$H:    &            &                &            &               \nl
SN 1972E & 0.040(0.065) & 0.056(0.006) & 0.143(0.014) & $-$0.103(0.066) \nl
SN 1980N & 0.182(0.054) & 0.021(0.002) & 0.054(0.005) & 0.128(0.054) \nl
SN 1983R & 0.070(0.089) & 0.074(0.007) & 0.189(0.019) & $-$0.119(0.091) \nl
SN 1999cp & 0.129(0.042) & 0.025(0.003) & 0.064(0.006) & 0.065(0.042) \nl
\nl
SN 1981B & $-$0.034(0.039) & 0.018(0.002) & 0.046(0.005) & $-$0.080(0.039) \nl
SN 1981D & 0.086(0.037) & 0.021(0.002) & 0.054(0.005) & 0.033(0.038) \nl
SN 1998bu & 0.961(0.056) & 0.024(0.002) & 0.061(0.006) & 0.899(0.057) \nl
SN 1999cl & 1.613(0.116) & 0.038(0.004) & 0.097(0.010) & 1.516(0.116) \nl
\nl

V$-$K:    &            &                &            &               \nl
SN 1972E & 0.094(0.065) & 0.056(0.006) & 0.154(0.015) & $-$0.061(0.067) \nl
SN 1980N & 0.120(0.033) & 0.021(0.002) & 0.058(0.006) & 0.062(0.033) \nl
SN 1983R & 0.255(0.099) & 0.074(0.007) & 0.204(0.020) & 0.051(0.101) \nl
SN 1999cp & 0.185(0.067) & 0.025(0.003) & 0.069(0.007) & 0.116(0.067) \nl
\nl
SN 1981B & 0.066(0.044) & 0.018(0.002) & 0.050(0.005) & 0.016(0.044) \nl
SN 1981D & $-$0.052(0.038) & 0.021(0.002) & 0.058(0.006) & $-$0.109(0.039) \nl
SN 1998bu & 1.010(0.041) & 0.024(0.002) & 0.066(0.007) & 0.944(0.042) \nl
SN 1999cl & 1.808(0.084) & 0.038(0.004) & 0.105(0.010) & 1.703(0.084) \nl

\nl
\enddata   
\tablenotetext{a}{All values in columns 2 through 5 are measured in
magnitudes. $^b$Galactic reddening from Schlegel et al. (1998).
The values in column 5 (implied host reddening) are equal to the 
values in column 2 (derived total reddening) minus
the values in column 4 (assumed Galactic reddening).}
\end{deluxetable}

\clearpage

\begin{deluxetable}{cccccc}
\tablewidth{0pc}
\tablecaption{Infrared Light Curve Templates}
\tablehead{
\colhead{$t$} & \colhead{M$_J$} & 
\colhead{M$_H$} &
\colhead{M$_K$} &
\colhead{M$_K$} &
\colhead{M$_K$} }
\startdata

$\Delta$: & 0.0 & 0.0 & $-$0.38 & 0.0 & +0.23  \nl
\nl 
$-$9.0 & $-$18.403 & $-$18.195 & $-$18.655 & $-$18.246 & $-$18.033 \nl
$-$8.0 & $-$18.540 & $-$18.293 & $-$18.768 & $-$18.359 & $-$18.146 \nl
$-$7.0 & $-$18.641 & $-$18.363 & $-$18.852 & $-$18.443 & $-$18.230 \nl
$-$6.0 & $-$18.733 & $-$18.432 & $-$18.937 & $-$18.528 & $-$18.315 \nl
$-$5.0 & $-$18.836 & $-$18.521 & $-$19.040 & $-$18.631 & $-$18.419 \nl

$-$4.0 & $-$18.859 & $-$18.538 & $-$19.072 & $-$18.663 & $-$18.450 \nl
$-$3.0 & $-$18.858 & $-$18.539 & $-$19.088 & $-$18.679 & $-$18.466 \nl
$-$2.0 & $-$18.835 & $-$18.528 & $-$19.092 & $-$18.683 & $-$18.465 \nl
$-$1.0 & $-$18.793 & $-$18.505 & $-$19.078 & $-$18.674 & $-$18.448 \nl
   0.0 & $-$18.730 & $-$18.469 & $-$19.050 & $-$18.654 & $-$18.420 \nl

   1.0 & $-$18.647 & $-$18.423 & $-$19.010 & $-$18.622 & $-$18.385 \nl
   2.0 & $-$18.546 & $-$18.367 & $-$18.966 & $-$18.581 & $-$18.343 \nl
   3.0 & $-$18.427 & $-$18.300 & $-$18.918 & $-$18.529 & $-$18.290 \nl
   4.0 & $-$18.291 & $-$18.226 & $-$18.867 & $-$18.470 & $-$18.224 \nl
   5.0 & $-$18.138 & $-$18.143 & $-$18.806 & $-$18.401 & $-$18.145 \nl

   6.0 & $-$17.969 & $-$18.170 & $-$18.738 & $-$18.326 & $-$18.059 \nl
   7.0 & $-$17.785 & $-$18.218 & $-$18.723 & $-$18.307 & $-$18.032 \nl
   8.0 & $-$17.587 & $-$18.260 & $-$18.756 & $-$18.334 & $-$18.053 \nl
   9.0 & $-$17.376 & $-$18.298 & $-$18.784 & $-$18.356 & $-$18.067 \nl
  10.0 & $-$17.278 & $-$18.331 & $-$18.811 & $-$18.373 & $-$18.075 \nl

  11.0 & $-$17.309 & $-$18.361 & $-$18.835 & $-$18.388 & $-$18.080 \nl
  12.0 & $-$17.337 & $-$18.388 & $-$18.854 & $-$18.399 & $-$18.080 \nl
  13.0 & $-$17.363 & $-$18.413 & $-$18.870 & $-$18.408 & $-$18.081 \nl
  14.0 & $-$17.389 & $-$18.437 & $-$18.886 & $-$18.417 & $-$18.080 \nl
  15.0 & $-$17.413 & $-$18.460 & $-$18.901 & $-$18.424 & $-$18.078 \nl

  16.0 & $-$17.436 & $-$18.482 & $-$18.915 & $-$18.430 & $-$18.074 \nl
  17.0 & $-$17.459 & $-$18.504 & $-$18.926 & $-$18.437 & $-$18.068 \nl
  18.0 & $-$17.482 & $-$18.526 & $-$18.933 & $-$18.443 & $-$18.064 \nl
  19.0 & $-$17.506 & $-$18.549 & $-$18.945 & $-$18.450 & $-$18.059 \nl
  20.0 & $-$17.530 & $-$18.572 & $-$18.961 & $-$18.458 & $-$18.057 \nl

  21.0 & $-$17.553 & $-$18.594 & $-$18.977 & $-$18.464 & $-$18.056 \nl
  22.0 & $-$17.578 & $-$18.618 & $-$18.995 & $-$18.472 & $-$18.057 \nl
  23.0 & $-$17.603 & $-$18.642 & $-$19.014 & $-$18.481 & $-$18.059 \nl
  24.0 & $-$17.628 & $-$18.666 & $-$19.031 & $-$18.489 & $-$18.060 \nl
  25.0 & $-$17.653 & $-$18.690 & $-$19.048 & $-$18.497 & $-$18.059 \nl

  26.0 & $-$17.679 & $-$18.715 & $-$19.065 & $-$18.507 & $-$18.058 \nl
  27.0 & $-$17.706 & $-$18.740 & $-$19.080 & $-$18.516 & $-$18.060 \nl
\nl

\enddata   
\tablenotetext{a}{Time $t$ is the number of days since B-band maximum.
These infrared templates were generated using the regression lines
given in Table 9, and assuming that the fiducial Type Ia SN has M$_V$ = 
$-$19.34 (Jha et al. 1999b).}
\end{deluxetable}

\begin{deluxetable}{cccccc}
\tablewidth{0pc}
\tablecaption{BVRI Transformation Coefficients$^a$}
\tablehead{
\colhead{Telescope} & \colhead{Applied to} & 
\colhead{$\epsilon _v$} & \colhead{$\mu _{bv}$} & 
\colhead{$\mu _{vr}$} &
\colhead{$\mu _{vi}$} }
\startdata

APO &  99aa      & +0.035(2) & 0.968(5) & 1.044(6) & 1.023(3) \nl 
APO &  99cl, 99cp & +0.038(2) & 0.967(3) & 1.057(6) & 1.020(2) \nl
MRO &  99cl, 99cp & +0.036(4) & 1.068(9) & 1.030(9) & 1.024(14) \nl
\nl
\enddata   
\tablenotetext{a}{These scale factors are used to multiply the
instrumental colors to produce V-band magnitudes (using instrumental
B$-$V), and B$-$V, V$-$R, and V$-$I colors, respectively.  The values
in parentheses are uncertainties in units of 0.001.}
\end{deluxetable}

%\end{document}

\begin{figure*}
\psfig{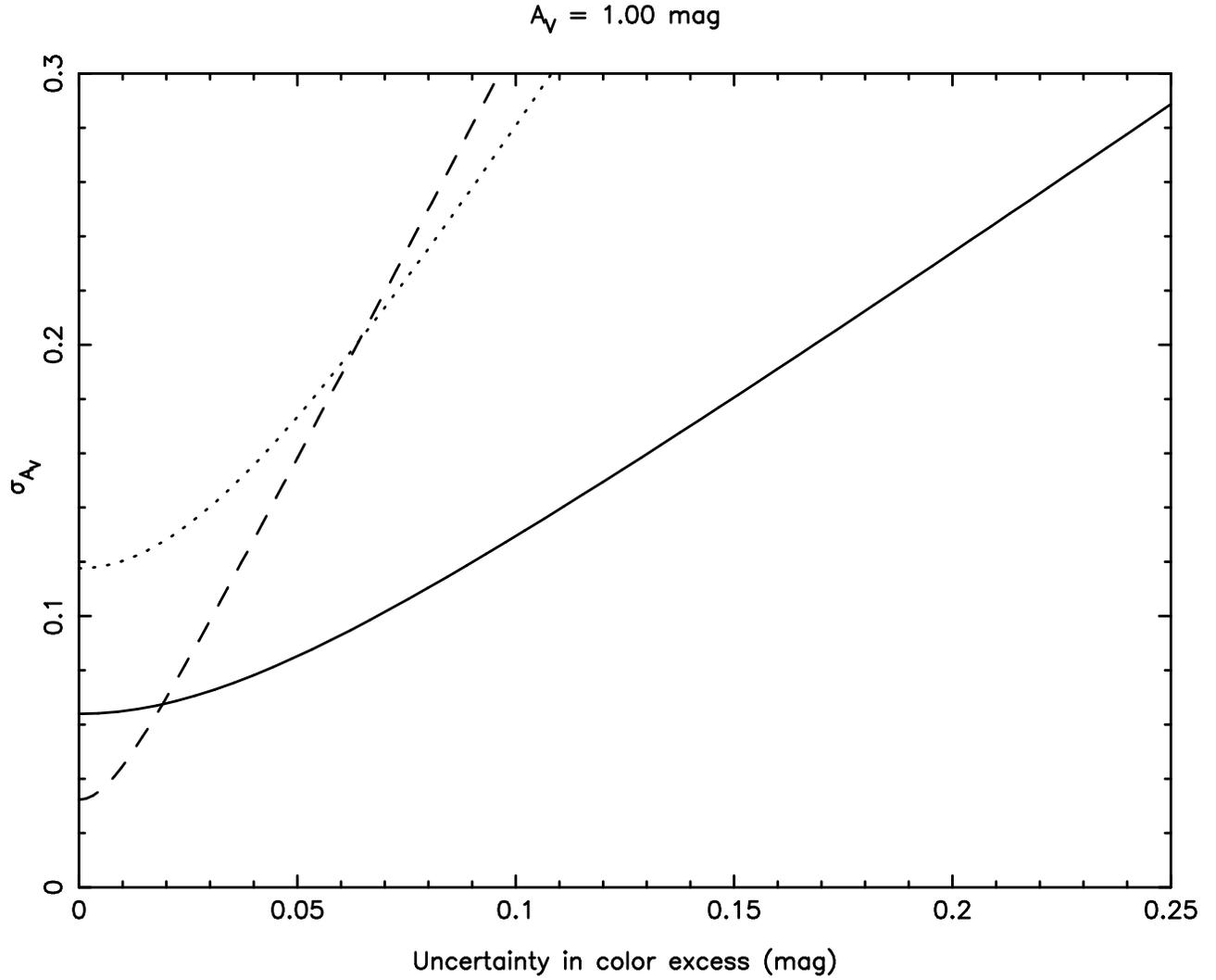} 
\caption{The uncertainty in the extinction A$_V$ as a function of the
uncertainty of the color excess, for A$_V$ = 1.00 mag.  Solid line: {\em
worst} case V$-$K scenario.  Dashed line: {\em best} case B$-$V scenario.  
Dotted line: most appropriate B$-$V case for Type Ia SNe. See text for
details.}
\end{figure*}

\begin{figure*}
\psfig{figure=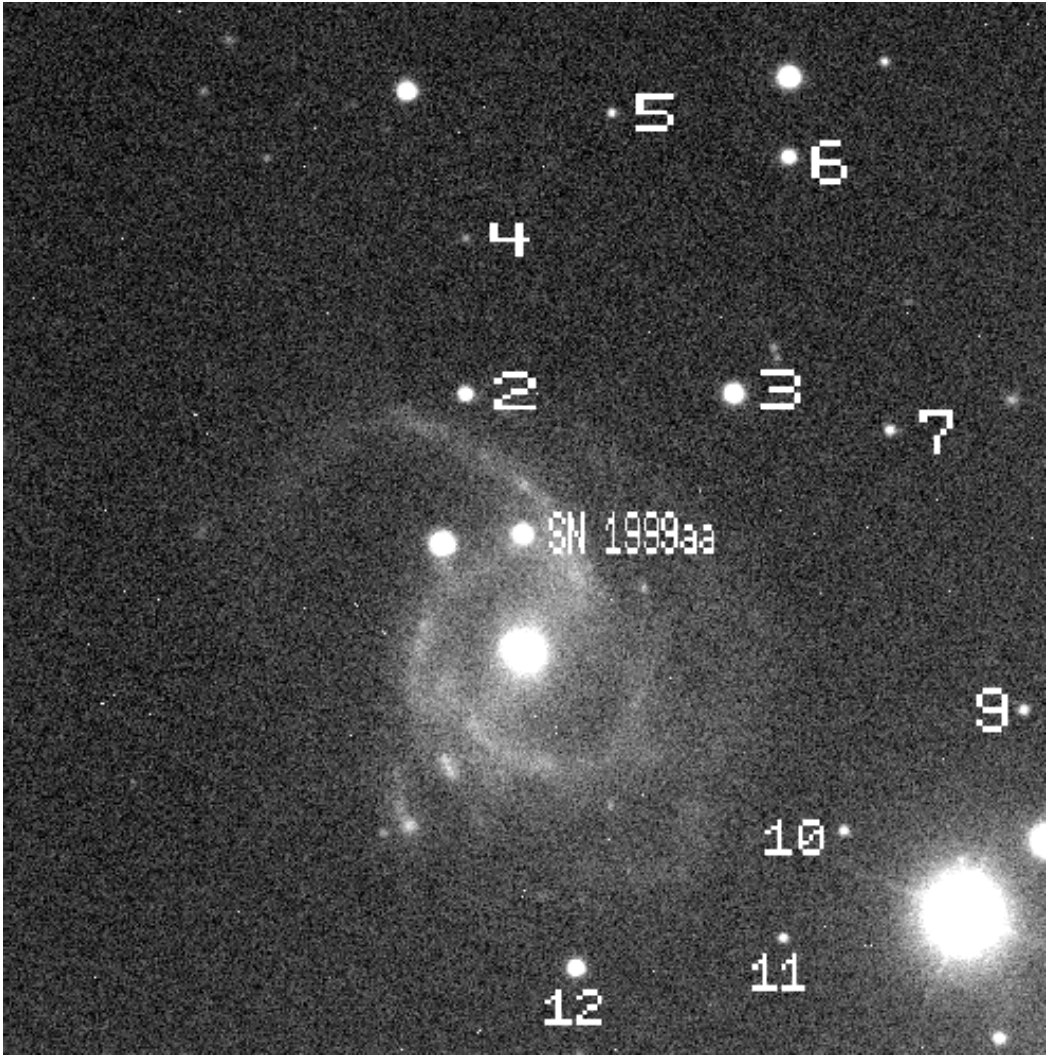,height=14cm}
\caption{A V-band image of NGC 2595 obtained at APO on 19 February 1999
UT, with SN 1999aa and the stars of the photometric sequence indicated.
The field is 4.8 arcmin on a side.  North is up, east to the left.}
\end{figure*}

\clearpage

\begin{figure*}
\psfig{figure=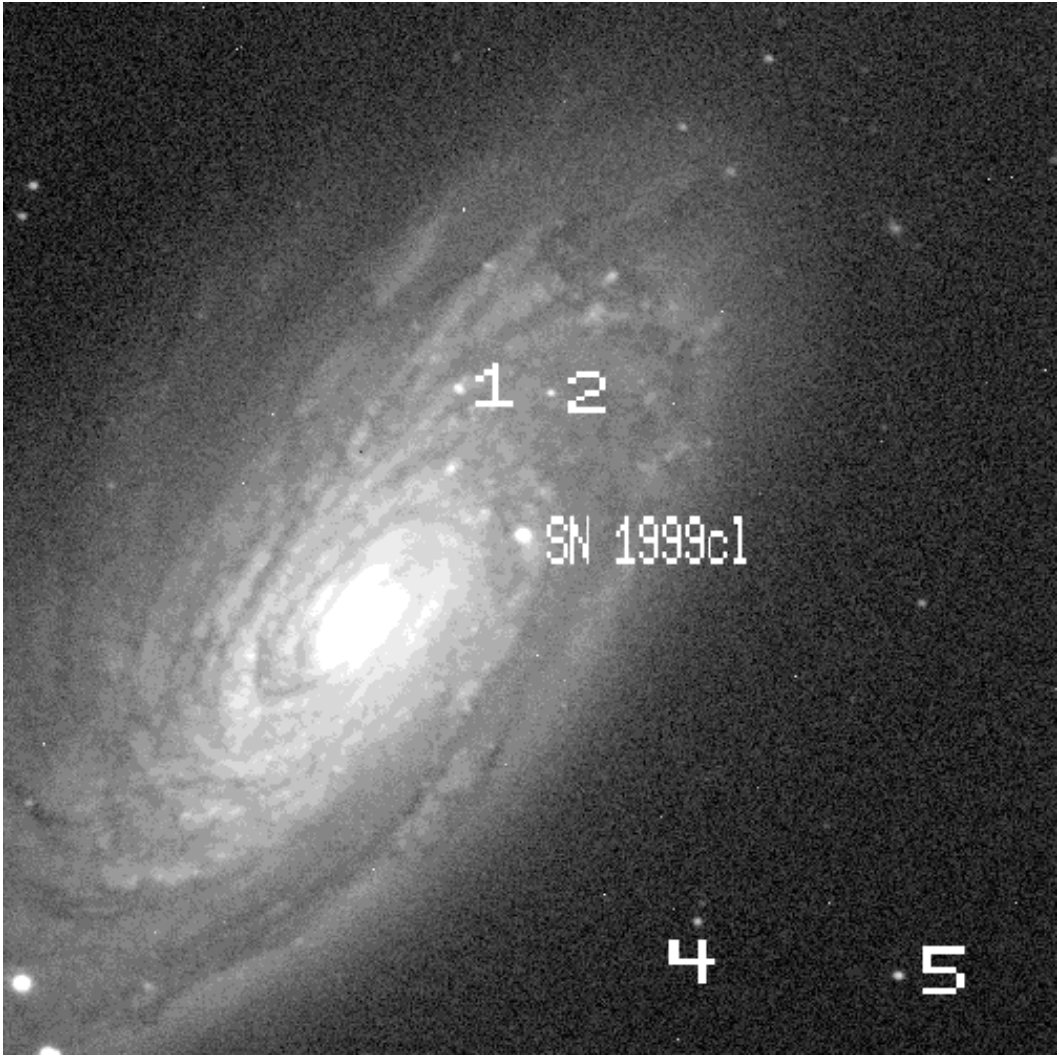,height=14cm}
\caption{A V-band image of NGC 4501 obtained with APO on 11 June 1999 UT,
with SN 1999cl and stars 1, 2, 4 and 5 of the photometric sequence
indicated.  The field is 4.8 arcmin on a side.  North is up, east to
the left.}
\end{figure*}

\begin{figure*}
\psfig{figure=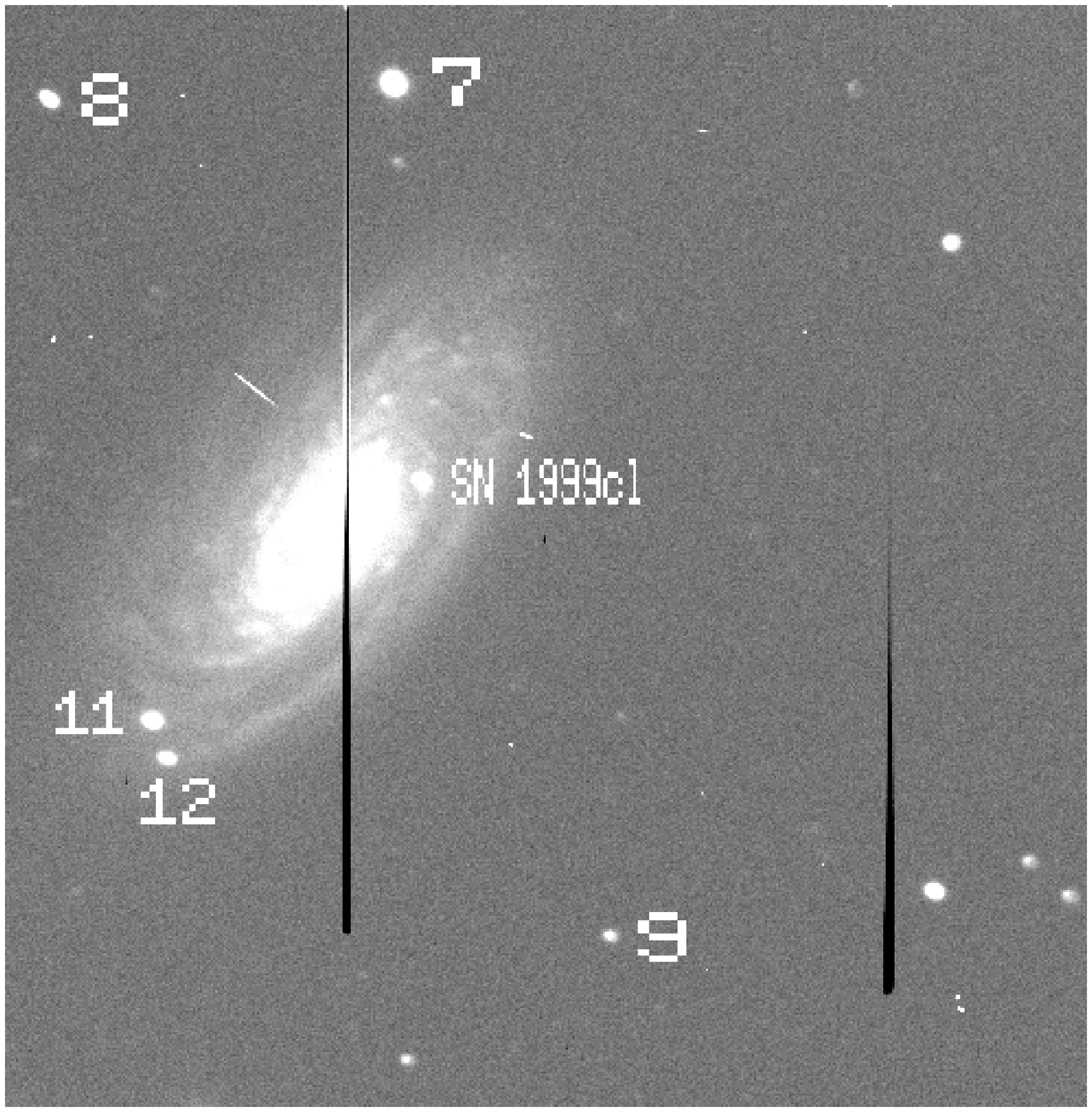,height=14cm}
\caption{A V-band image of NGC 4501 obtained with MRO on 6 July 1999 UT,
with SN 1999cl and other stars of the photometric sequence
indicated. The field is 11 arcmin on a side.  North is up, east to the
left.}
\end{figure*}

\begin{figure*}
\psfig{figure=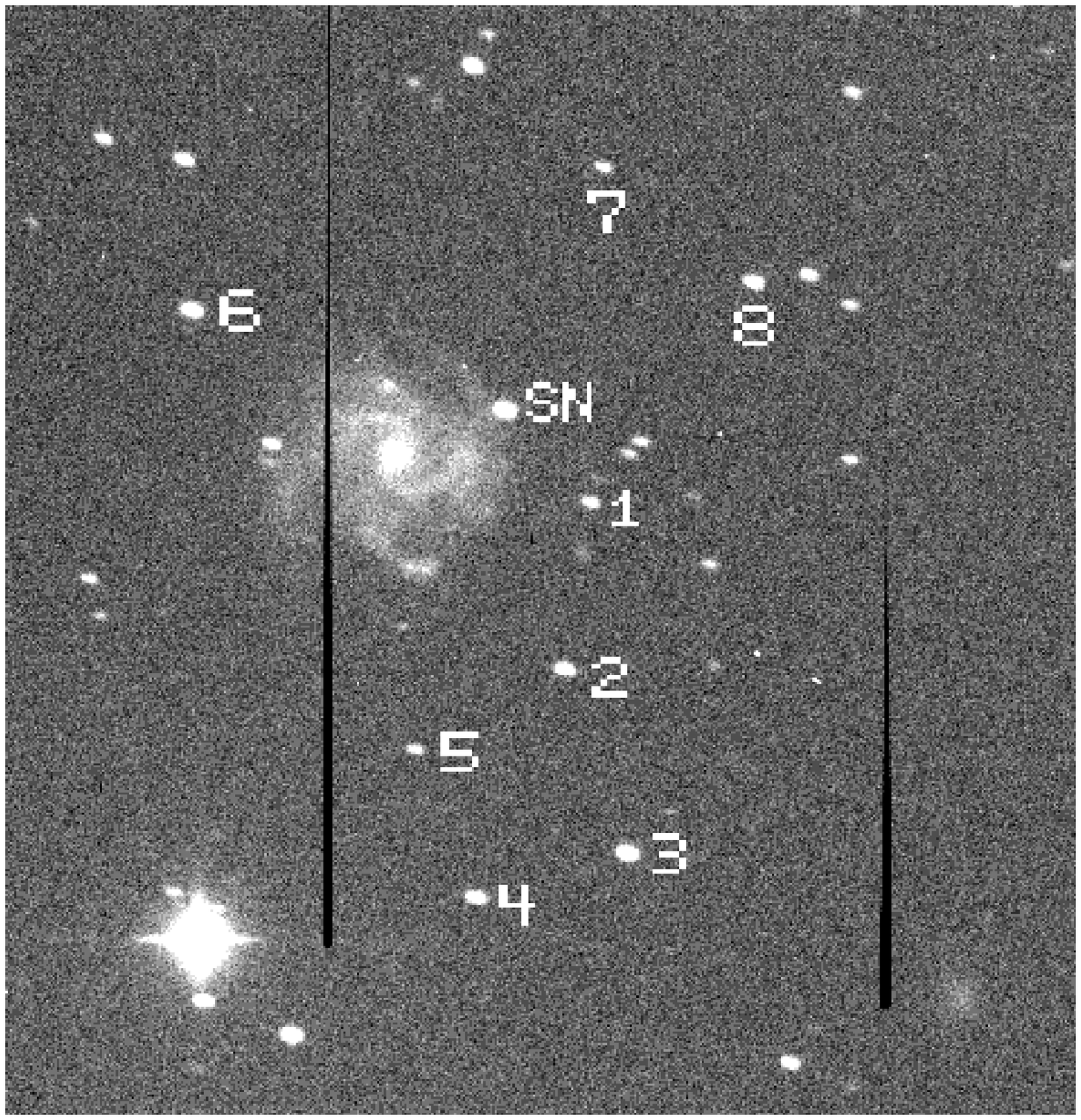,height=14cm}
\caption{A V-band image of NGC 5468 obtained with MRO on 9 July 1999 UT,   
with SN 1999cp and the stars of the photometric sequence indicated.
The field is 11 arcmin on a side.  North is up, east to the left.}
\end{figure*}

\begin{figure*}
\psfig{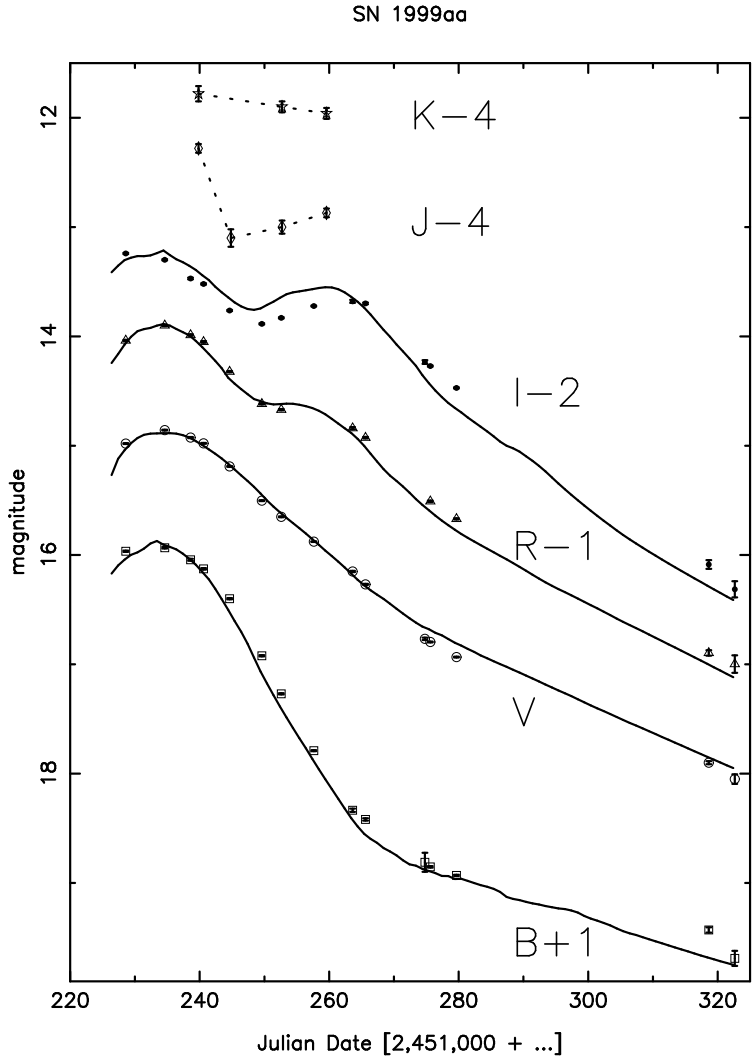}
\caption{BVRIJK photometry of SN 1999aa.  The B, R, I, J, and K
data have been offset vertically by +1, $-$1, $-$2, $-$4, and $-4$
magnitudes,
respectively.  The solid lines are based on MLCS empirical fits with
$\Delta$ = $-$0.47 mag.}
\end{figure*}

\begin{figure*}
\psfig{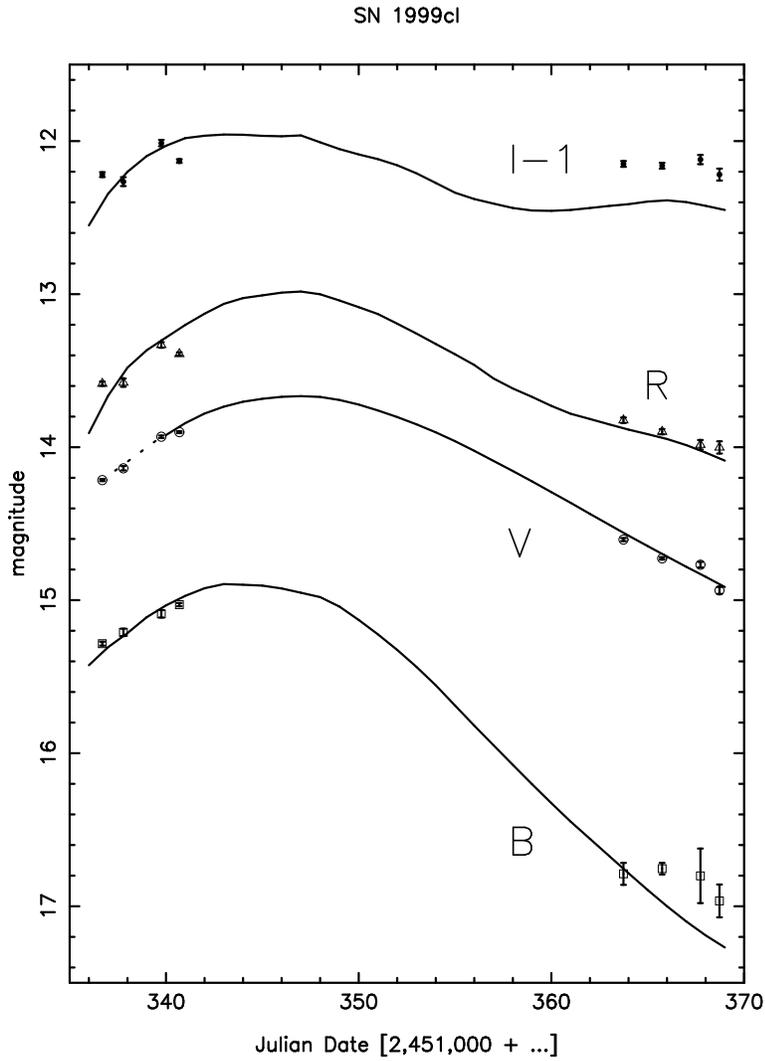}
\caption{BVRI photometry of SN 1999cl.  Only the I-band points have
been offset (by $-$1 magnitude).  Also shown are the MLCS empirical fits
with $\Delta$ = +0.20 mag.  For the purposes of determining
V {\em minus} infrared colors at early times, the dotted line fit
to the early V-band points will be used.}
\end{figure*}

\begin{figure*}
\psfig{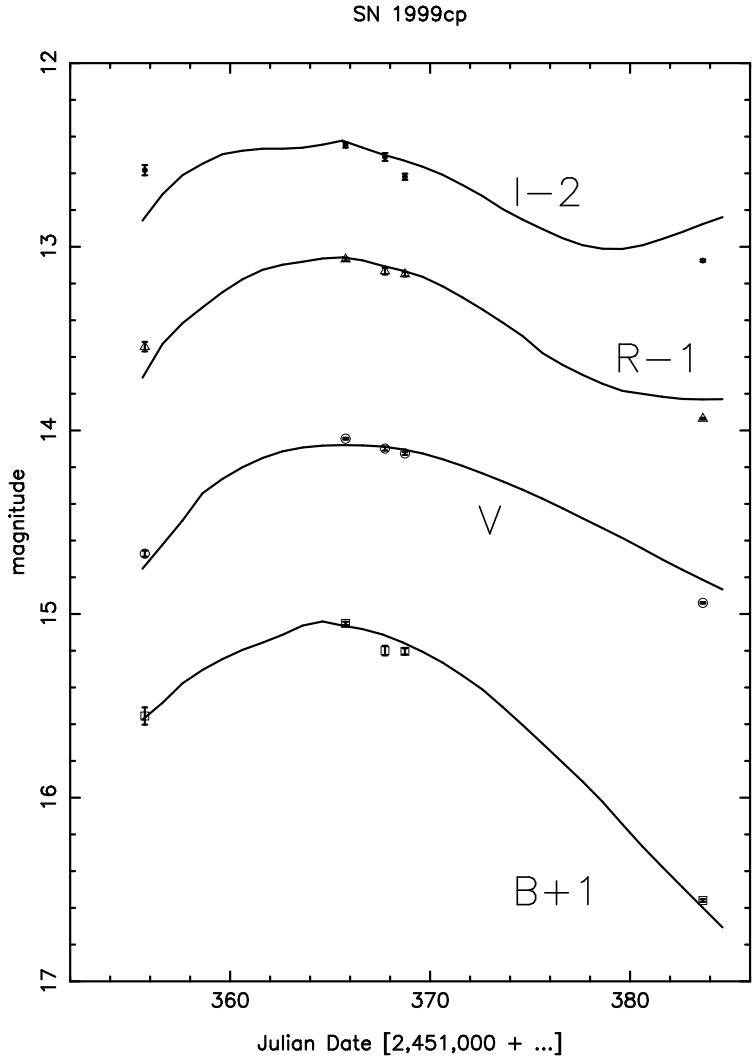}
\caption{BVRI photometry of SN 1999cp.  The B, R, and I
data have been offset vertically by +1, $-$1, and $-$2 magnitudes,
respectively. The solid lines are based on MLCS empirical fits
with $\Delta$ = $-$0.31 mag.}
\end{figure*}
 
\clearpage

\begin{figure*}
\psfig{figure=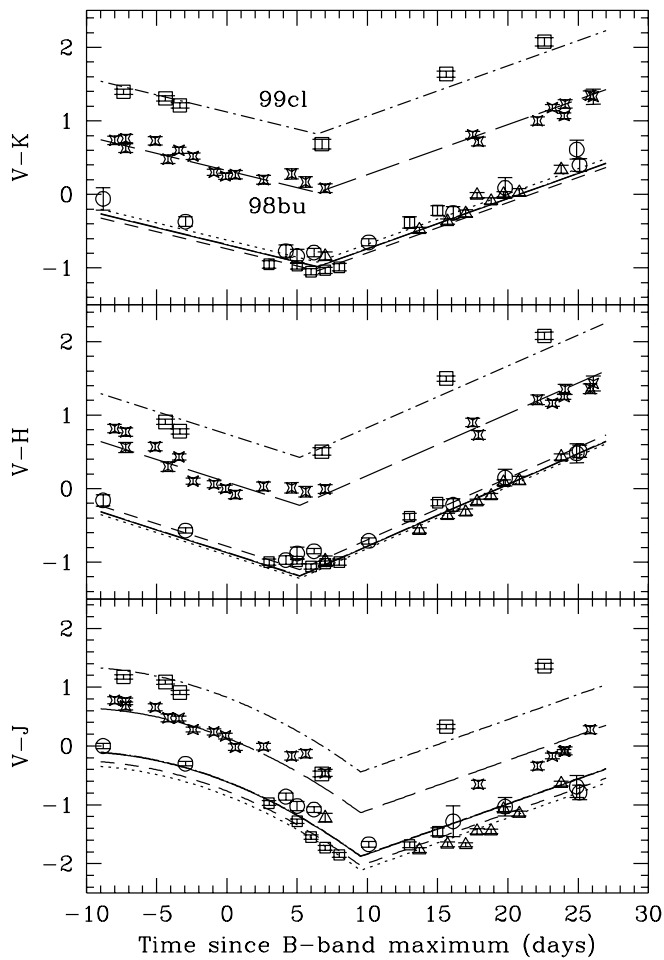,height=14cm}
\caption{Observed V {\em minus} near infrared colors of Type Ia
SNe.  Data for SNe 1972E, 1980N, 1983R, and 1999cp are represented
by open circles.  Other
points represent specific objects: SN 1981B (triangles), SN 1981D
(smaller open squares), SN 1998bu (four pointed stars), and SN 1999cl
(larger open squares).  The solid lines represent the fiducial (i.e.
unreddened) loci.  The dotted lines are fits for SN 1981B, while the short
dashed lines are for SN 1981D, the long dashed lines are for SN 1998bu,
and the dot-dashed lines are for SN 1999cl.  The slopes of the fits
are determined by the entire data set for a given color.}
\end{figure*}

\begin{figure*}
\psfig{figure=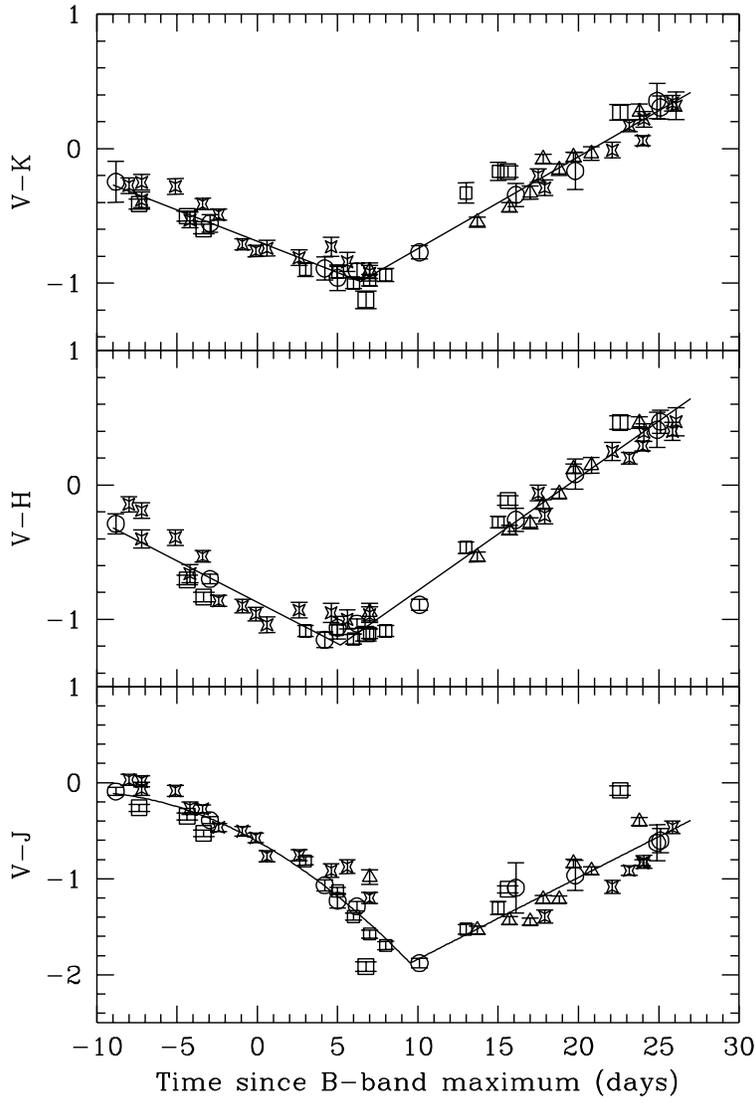,height=16cm}
\caption{The data shown in Fig. 9 have been adjusted for the offsets
a$_i$ in Table 10.  This shows the scatter of the data
which would be observed if all the SNe under study were reduced 
to the same loci.  The symbols are the same as in Fig. 9.}
\end{figure*}

\begin{figure*}
\psfig{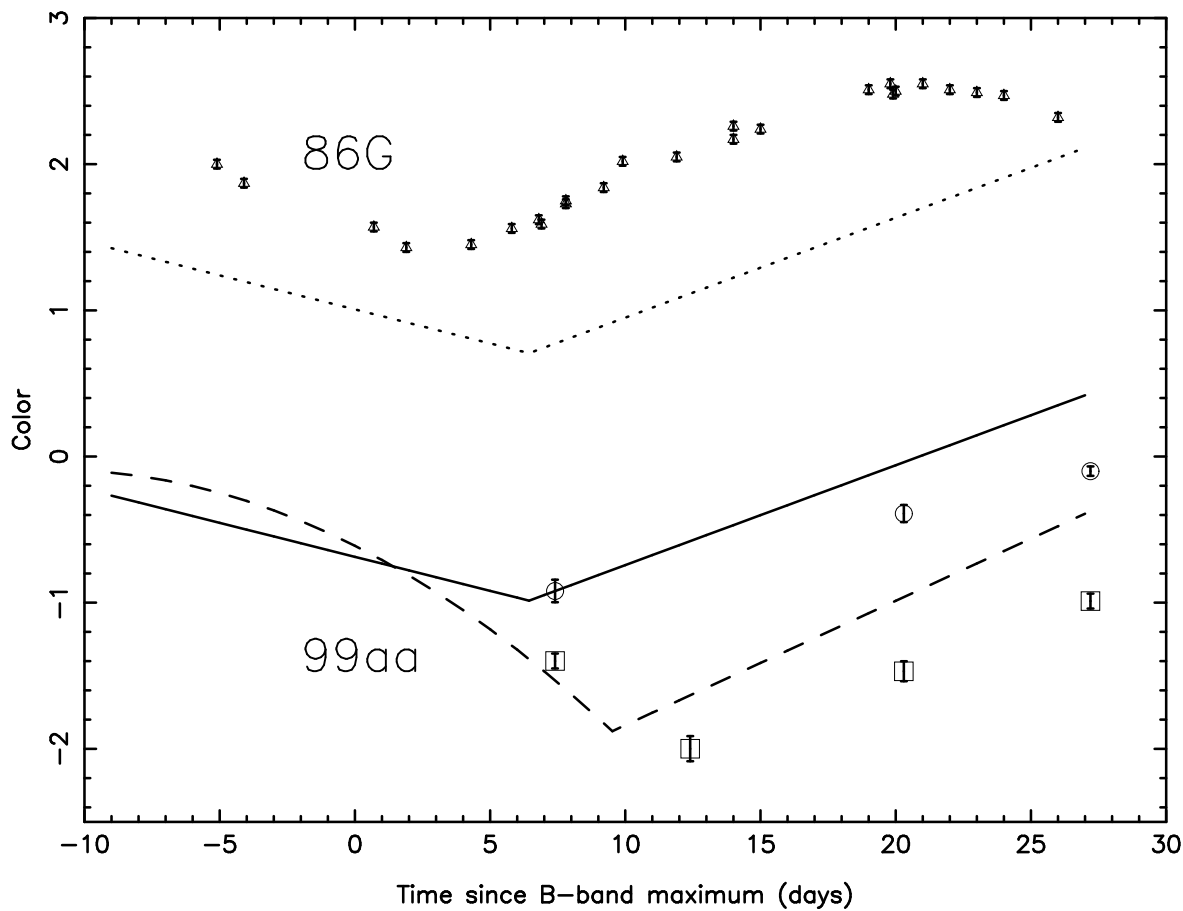}
\caption{V$-$J (open squares) and V$-$K data (open circles) for SN 1999aa,
corrected for Galactic reddening.  V$-$K colors for SN 1986G (triangles)
are taken from Frogel et al. (1987).  The solid line is the unreddened
V$-$K locus, while the dashed line is the unreddened V$-$J locus. The
dotted line is the unreddened V$-$K locus offset by 3.1 $\times$ 0.888
$\times$ 0.615 = 1.693 mag, the expected V$-$K excess for Galactic
reddening parameters and the color excess E(B$-$V) = 0.615 mag given by
Phillips et al. (1999).  Clearly, the color loci based on Type Ia SNe with
$-0.38 \leq \Delta \leq +0.23$ mag do {\em not} fit the data for
overluminous, slow decliners such as SN 1999aa, or underluminous, fast
decliners such as SN 1986G.}
\end{figure*}

\clearpage

\begin{figure*} 
\psfig{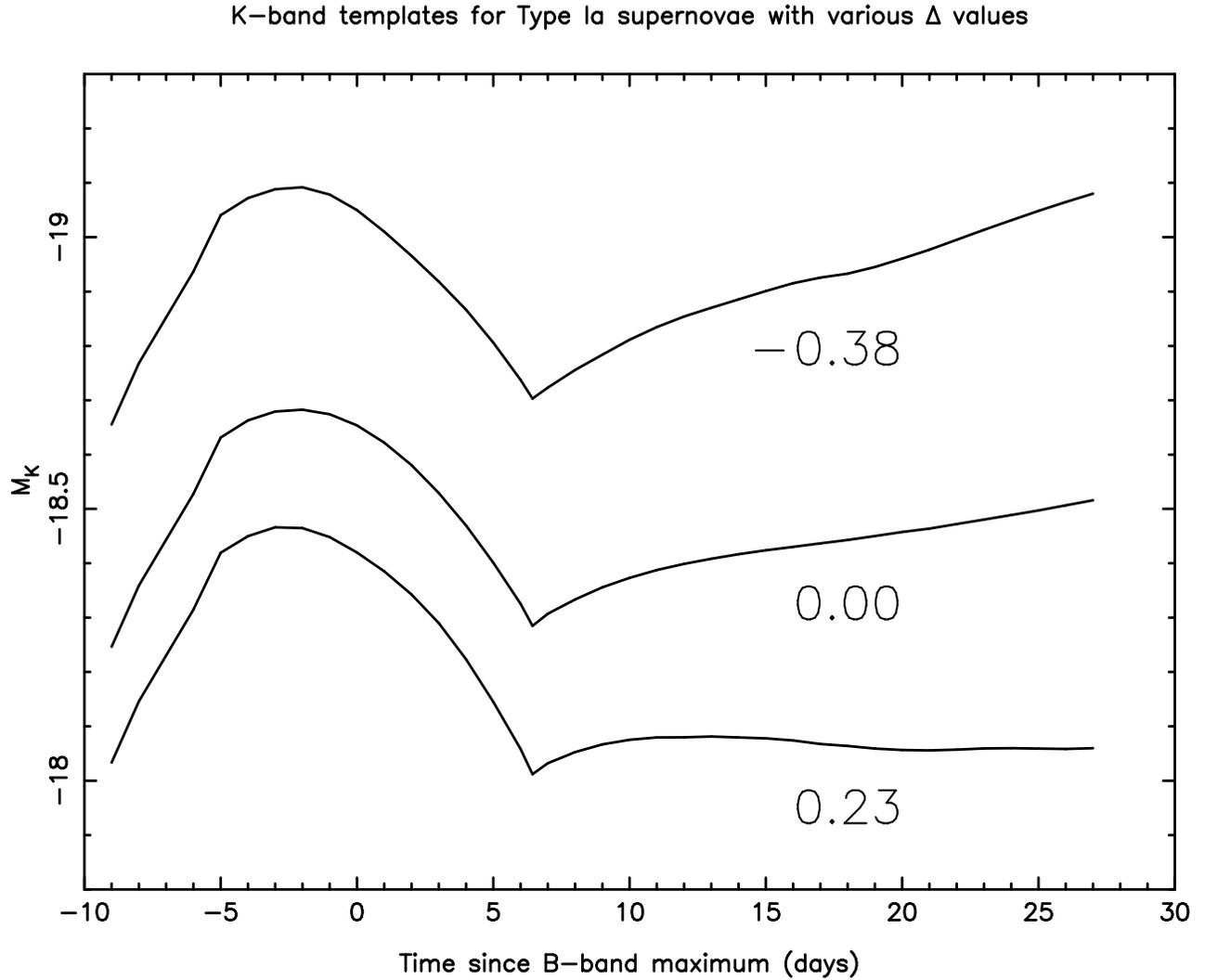} 
\caption{Since K = V $-$ (V$-$K), the existence of V-band light curve
templates and the V$-$K relationship from Table 9 and Fig. 10 predict the
K-band templates given here.  Given the uncertainty in the absolute
magnitude zero points and the zero point of our unreddened V$-$K locus,
these loci might be shifted $\pm$ 0.2 mag, keeping their relative
separations constant.}
\end{figure*}

\begin{figure*} 
\psfig{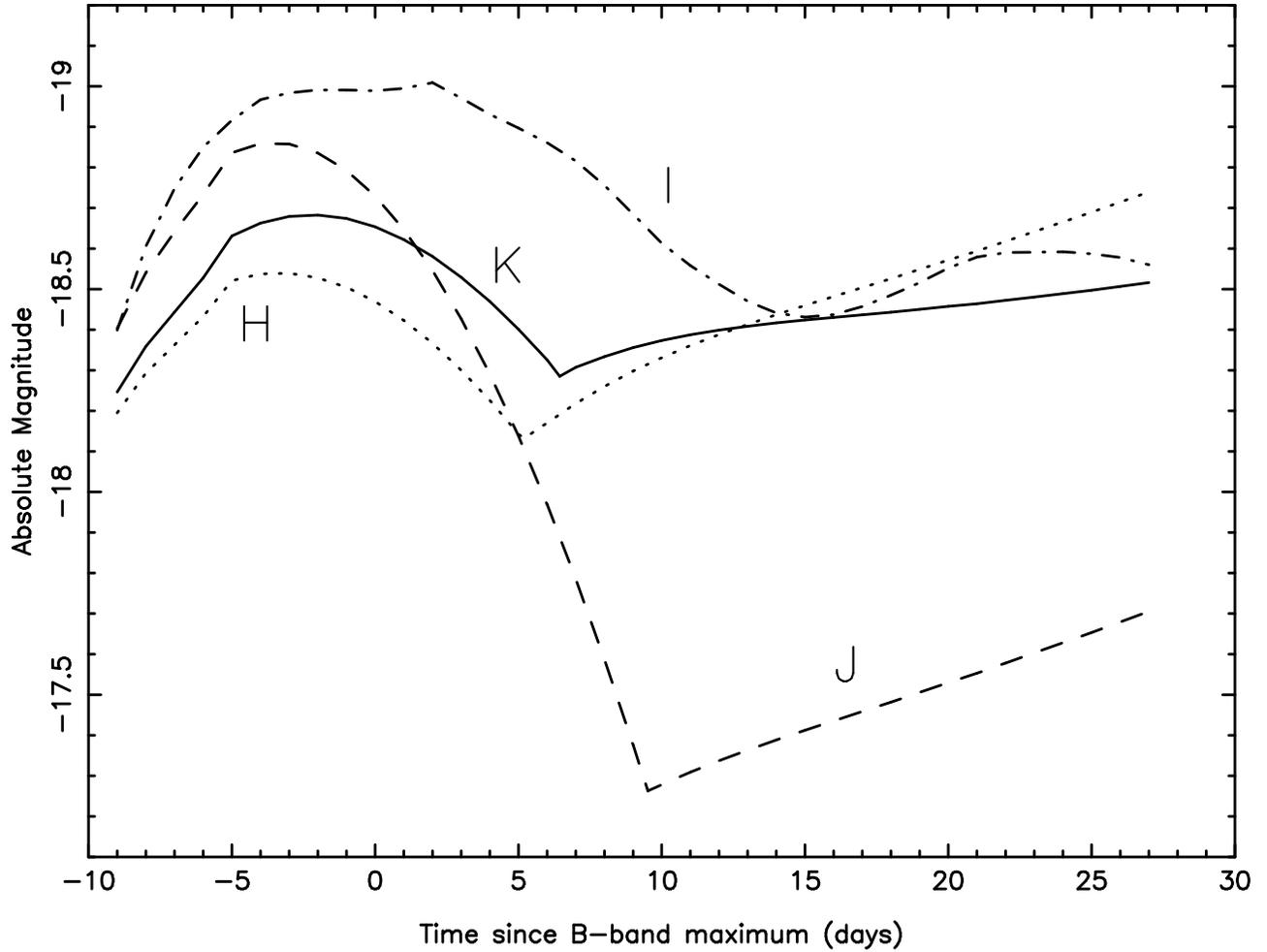} 
\caption{Near infrared light curve templates for Type Ia SNe with
$\Delta$ = 0.0 mag.  I = dot-dashed line.
J = dashed line; H = dotted line; K = solid line.
As with Fig. 12, these loci are uncertain to $\pm$ 0.2 mag.} 
\end{figure*}

\end{document}